\documentclass[%
 reprint,
 amsmath,amssymb,
 aps,
prb,
]{revtex4-1}

\usepackage{graphicx}
\usepackage{dcolumn}
\usepackage{bm}
\usepackage{hyperref}

\usepackage[T1]{fontenc}
\DeclareUnicodeCharacter{00A0}{~}
\usepackage{etoolbox}
\usepackage{psfrag}
\usepackage{ae}
\usepackage{amsfonts}
\usepackage{siunitx}
\usepackage{color}
\usepackage{newunicodechar}
\usepackage{dcolumn}
\usepackage{comment}

\newcommand{\bnabla} {\mbox{\boldmath$\nabla$}}

\usepackage{epstopdf}

\definecolor{cream}{RGB}{222,217,201}

\begin{document}

\preprint{APS/123-QED}

\title{Continuum theory of electrostatic-elastic coupling interactions in colloidal crystals}

\author{Hao Wu}
 \email{wuhao@ucas.ac.cn}
\affiliation{%
 Wenzhou Institute, University of Chinese Academy of Sciences, Wenzhou, Zhejiang 325001, China.
}%

\author{Zhong-Can Ou-Yang}
\affiliation{%
 Institute of Theoretical Physics, Chinese Academy of Sciences, Beijing 100190, China;
}

\author{Rudolf Podgornik}
 \email{podgornikrudolf@ucas.ac.cn}
 \affiliation{%
 Wenzhou Institute, University of Chinese Academy of Sciences, Wenzhou, Zhejiang 325001, China.
}%
\affiliation{%
 School of Physical Sciences, University of Chinese Academy of Sciences, Beijing 100049, China.
}%
\affiliation{
 Kavli Institute for Theoretical Sciences, University of Chinese Academy of Sciences, Beijing 100049, China.
}%
\affiliation{
 Institute of Physics, Chinese Academy of Sciences, Beijing 100190, China.
}%


\date{\today}

\begin{abstract}
A mobile Coulomb gas permeating a fixed background crystalline lattice of charged colloidal crystals is subject to an  { electrostatic-elastic} coupling, which we study on the continuum level by introducing a minimal coupling between electrostatic and displacement fields. We derive linearized, Debye-H{\"u}ckel-like mean-field equations that can be analytically solved, incorporating the minimal coupling between electrostatic and displacement fields leading to an additional eﬀective attractive interaction between mobile charges, that depends in general on the strength of the coupling between the electrostatic and displacement fields. By analyzing the Gaussian fluctuations around the mean-field solution we also identify and quantify the region of its stability in terms of the  { electrostatic-elastic} screening length. This detailed continuum theory incorporating the standard lattice elasticity and electrostatics of mobile charges provides a baseline to investigate the  { electrostatic-elastic} coupling for microscopic models in colloid science and materials science.

\end{abstract}

\maketitle


\section{\label{sec:intro}Introduction}

Colloidal crystals present an interesting model of condensed matter systems where atoms and electrons are substituted by charged colloidal macroions and mobile, small electrolyte ions \cite{Olvera2019}. They can form in systems with purely repulsive long-range interactions \cite{WILLIAMS1974225,Blaaderen,PhysRevLett.62.1524,PhysRevE.91.032310} or in systems with combined short-range attraction and long-range repulsion \cite{Dotera2014}.
The Deryaguin-Landau-Verwey-Overbeek (DLVO) type interactions between these constituents can be modified and engineered \cite{Kegel}, yielding valuable insights not only into the properties of colloidal crystals themselves, but illuminating also the theoretical foundations of atomic and molecular materials such as crystalline elasticity, crystal melting, lattice defects and dynamics, to name just a few \cite{lechner2009defect,lechner2009point,Olvera2019}. Colloids of various components and different size/charge asymmetries have been assembled into diverse crystalline structures and have served as experimental models to study phase behaviors and self-assembly processes \cite{DINSMORE19985,Olvera2019}. Of particular interest are the binary colloidal crystals where the smaller, mobile component occupies interstitial space in the lattice, formed by the larger colloid component, and can exhibit a {\sl localized-to-delocalized transition} \cite{Schmitz1999,lin2022superionic}, as is the case in size-asymmetric DNA-functionalized nanoparticles \cite{Olvera2019}. A similar type of behavior is also known to occur in a two-dimensional (2D) crystalline array of cylindrical macroions  with interstitial linear polyelectrolyte chains exhibiting a behavior analogous to that of electrons in a 2D crystal \cite{Saslow2005}. In the context of colloidal crystals, this localized-to-delocalized transition has been dubbed the {\sl ionic to metallic transition} \cite{lin2022superionic}.

Attractive interactions between like-charged colloidal particles that cannot be rationalized within the DLVO paradigm \cite{Mar21}, first seen in detailed electrostatic double-layer simulations 
{\cite{BOROUDJERDI2005129}}, have been directly observed in various experiments \cite{Borkovec20181,LUDWIG2020137} and have been studied extensively in the framework of different theoretical approaches 
\cite{Perspective}, however, without any clearcut consensus emerging as to their underlying universal mechanism(s).  
This {\sl embarras de richesses} \cite{podgornik2021embarras} in theoretical understanding of the non-DLVO colloid interactions stems partially from different microscopic models and partially from different approximations applied to solve these models. Among the variation on the model theme we can refer to the most recent proposal of charge regulation at the interacting colloid surfaces yielding an accurate description of the observed force profiles, including their attractive part,  down to a few nm \cite{TREFALT20179,Borkovec20181,Borkovec20182}. On the other hand, the departure from the mean-field approximation is still best rationalized in terms of the {\sl weak-strong coupling dichotomy} of confined Coulomb fluids that yields an attractive electrostatic interaction between identical colloids \cite{BOROUDJERDI2005129,Perspective}.

\begin{figure*}[t!]
\includegraphics[width=16cm]{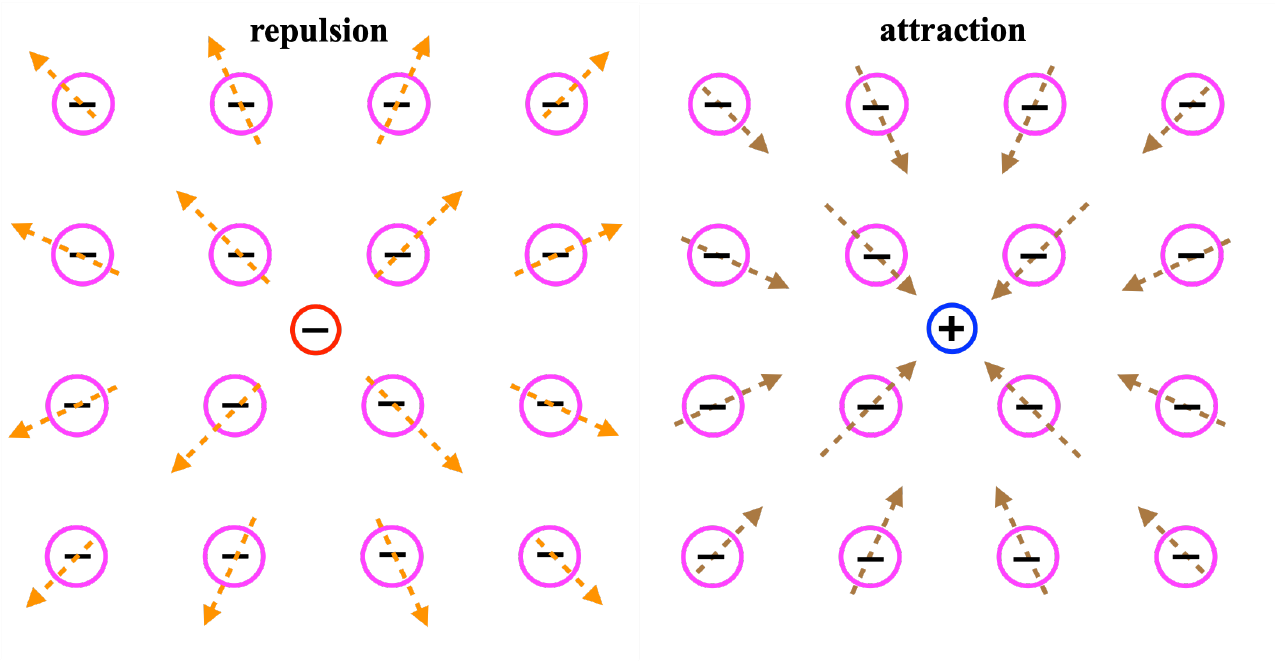}
\caption{A colloidal lattice {  contains} an interstitial {  charged} "intrusion". {  A negative (red circle) or positive (blue circle) interstitial test charge imposes a local elastic dilation (orange arrow) or contraction (brown arrow)} on the ideal colloidal crystal lattice negatively charged {  (magenta circle)}. The interstitial charge therefore induces a local change in volume/density of the lattice.}
\label{Fig0}
\end{figure*}

As a consequence, discovering new phenomena that would display some novel features of the remaining {\sl unexamined life} of the anomalous attractive forces in charged colloids are thus certainly worth to focus on more in detail \cite{LUDWIG2020137}. 
Elastic deformation effects  are ubiquitous in crystalline solids \cite{Sutton} where various defects, such as dislocations, interstitials and vacancies, create elastic displacement fields that allow them to interact with long-range interactions \cite{lechner2009defect, lechner2009point, lechner2013self,eisenmann2005pair,he2013interaction}. These lattice Hookian elasticity-mediated forces between point-defects have been shown to be attractive, leading in the case of interstitials and vacancies to the formation of defect strings despite the purely repulsive forces between the colloidal particles \cite{lechner2009defect, lechner2009point, pertsinidis2001equilibrium, lechner2014role}. While the phenomenology of these lattice-mediated attractions  bears some similarity to the anomalous attractions between identically charged colloids (see above), the underlying mechanism seems to be quite different and involves an additional (elastic) degree of freedom. Similar lattice-mediated attractions can be surmised to exist between diffusing test charges in the interstitial mobile Coulomb fluid but modified by the ubiquitous electrostatic interactions. This situation is superficially reminiscent of the phonon-mediated attractions between electrons in superconductor Cooper pairing \cite{bardeen1957theory}, but of course lacks its quantum basis, {  therefore could be called "thermal Cooper pairing" \cite{wu2024electrostatic}}. Motivated by these findings, we will attempt to formulate {  the} simplest possible continuum model of the  { electrostatic-elastic} coupling in a model colloidal crystal and ascertain to what extent it mimics the behavior of other solid state systems.

In what follows we will present a mean-field continuum model of a background 3D crystalline lattice with an interstitial Coulomb fluid and calculate explicitly the effective interaction between two test interstitial charges. The model is based on continuum elasticity of the crystalline lattice, 
and Poisson-Boltzmann {  (PB)} mean-field electrostatics for the interstitial small mobile ion component. In addition,  we will assume a minimal {{ electrostatic-elastic}} coupling due to the colloidal lattice-mobile ion interactions which implies a local lattice dilation/constriction at the  location of an interstitial mobile charge. In thermodynamic equilibrium this coupling will lead to two equilibrium equations: a modified {\sl Navier equation} for the elastic displacement and a modified {\sl {  PB} equation} for the electrostatic field. Linearizing and solving these two equations explicitly for a test point  interstitial charge, we derive an effective interaction between two interstitial test sources that displays a variable attractive component depending on the strength of the bare electrostatic screening as well as the  { electrostatic-elastic} coupling. Notably and contrary to the anomalous attraction in other colloidal systems (see above) the effective attraction emerges already on the mean-field level, the reason being that the  { electrostatic-elastic} coupling leads to a two-field theory and thus straightforwardly implies a decrease of the total interaction free energy.

\section{Minimal coupling model: mean-field theory}

We consider a minimal model describing the  { electrostatic-elastic} coupling as follows: we have an ideal colloid crystalline lattice of charged macroions, see Fig. \ref{Fig0}, embedded in a bathing Coulomb fluid of mobile univalent positive and negative ions, in thermal equilibrium with a bulk phase. In the spirit of a continuum theory both of the components, the mobile ions as well as the lattice colloids, are assumed to be point-like. In addition we assume that the lattice colloids can exhibit only small deviations from their equilibrium lattice positions, while the bathing Coulomb fluid ions are fully mobile.   

Within this model we assume that the free energy of the  system as being decomposable into an elastic part, describing the lattice colloids, an electrostatic part,  describing the mobile Coulomb fluid, as well as a minimal coupling part that connects the electrostatic field with elastic deformation. 
Our aim is then to calculate an effective  { electrostatic-elastic} interaction between two point-like interstitial test particles ("intrusions") immersed into the system, that interact with the Coulomb fluid components electrostatically as well as with the background colloidal lattice elastically.  

The electrostatic free energy of the Coulomb fluid is furthermore assumed to coincide with the standard {  PB} free energy \cite{Mar21}, while the elastic energy of the colloidal lattice is of a standard Hookeian form, allowing a further simplification if one assumes that the elastic medium is isotropic \cite{Sutton}. The elastic constants can be expressed as lattice sums of the interaction potential between colloid macroions in an undistorted lattice \cite{lechner2009point}, which is furthermore assumed to be three dimensional  \cite{QIanXie1994}. Finally, the  { electrostatic-elastic} coupling term embodies the effect that the interstitial charges have on the dilation (or contraction) of the background colloidal lattice. It features in two different contexts: the first is the coupling between the Coulomb fluid charge density and the elastic displacement, the second one is the coupling between the test charge particles and the neighboring background colloidal lattice. {  The two characteristics both have been included in our recent letters \cite{wu2024electrostatic}}.

{ The electrostatic free energy is based on the mean-field PB  description and is composed of an electrostatic energy and an ideal gas entropy of the mobile ions.} It can be written in different equivalent forms either in the form of a density functional theory~\cite{Lowen1993} or a field theory~\cite{maggs2016general}. We choose here a hybrid description of the form~\cite{Mar21}
\begin{eqnarray}
&& {\cal F}_{ES}[n_{\pm}, \phi ] = \int_V {\rm d}^3{\bm x}~ \Big[ - {\textstyle\frac12} \varepsilon \left(\bnabla \phi\right)^2 + e_0(n_+-n_- ) \phi + \nonumber\\
&&~~~~~+ k_{\rm B}T \sum_{\pm}\left( n_i \ln{n_i a^3} - n_i\right) - \sum_{\pm} \mu_i n_i + e_0 \rho_e \phi\Big]
\label{FES1}
\end{eqnarray}
where $\varepsilon$ is proportional to the dielectric permittivity, $e_0$ is the elementary charge,  $n_{\pm}$ are the densities of the diffusing mobile ions, $\mu_{\pm} = \mu_0$ are their chemical potentials which are identical since the mobile components are electroneutral in the bulk, $\phi$ is the mean electrostatic potential, while $\rho_e$ here and below corresponds to the source term pertaining to the point-like test charge source at the origin, with the density $\rho_e = \delta({{\bm x}})$. {Just as in the case of the (linearized) PB theory the test charge will allow us to calculate the electrostatic potential and the charge-charge interaction.} The minimization of this free energy yields the PB equation and the chemical potentials are expressed by the concentrations of the ions in the bulk.

For the elastic part we first of all rewrite the standard Hookeian elastic free energy for an isotropic elastic medium that is usually written in the form \cite{Sutton}
\begin{eqnarray}
{\cal F}_{EL}[u_{ik}] &=& \int_{V}{\rm d}^3{\bm x}~ \Big[ {\textstyle\frac 12} \lambda ({\rm Tr}~ u_{ik})^2  + \mu ~{\rm Tr}~u_{ik}^2 - f_i u_i\Big],~~~
\label{equela0}
\end{eqnarray}
where $u_i$ is the elastic displacement vector, $f_i$ is the external force density, $u_{ik}$ is the symmetrized deformation tensor, while $\lambda$ and $\mu$ are Lam{\' e} coefficients, which can be expressed as lattice sums of the  undistorted lattice \cite{lechner2009point}. The last term is the source term,  pertaining to the point-like interstitial elastic test  source at the origin. The source term in Eq. (\ref{FES1}) and the source term in Eq. (\ref{equela0}) both correspond to a test particle at the origin, but the former describes its electrostatic interaction and the latter describes its elastic interaction with the system.

The elastic constants can be obtained straightforwardly in the absence of thermal effects, starting from the total energy of the colloidal crystal \cite{lechner2009point}
\begin{equation}
   E = {\textstyle\frac12} \sum_{i\neq j}  \upsilon(\vert r_i - r_j\vert),
   \label{collint}
\end{equation}
where $\upsilon(\vert r_i - r_j\vert)$ is the pair potential between colloid particles  $i$ and $j$ in the undistorted lattice. Its form is often assumed to coincide with the Debye-H{\"u}ckel {  (DH)} interaction with an effective charge of the colloid \cite{WILLIAMS1974225,Blaaderen,PhysRevLett.62.1524,PhysRevE.91.032310}, that includes the effects of distorted double layers at high macroion volume fractions, but it should also include the details of the dissociation process, which are more difficult to incorporate.  
From here, in the case of an isotropic elastic material, one can derive the following identities for 
the Lam{\' e} elastic constants $\lambda$ and $\mu$ \cite{Schwerdtfeger2021,ashcroft1976}
\begin{eqnarray}
\lambda + \frac{2}{3} \mu &=& \frac{\rho}{18} {\sum_{i}}'  \Big( \upsilon''(r_i) r_i^2 - 2 \upsilon'(r_i)r_i\Big) \nonumber\\
\mu &=& \frac{\rho}{30}{\sum_{i}}'  \Big( \upsilon''(r_i) r_i^2 + 4\upsilon'(r_i)r_i \Big),
\label{consts}
\end{eqnarray}
where $\upsilon'(r)$ and $\upsilon''(r)$ are the first and second derivative of the pair potential, respectively, $\rho$ is the colloid number density, $r_i$ is the distance
of particle $i$ from the origin. In the summation, the floating prime indicates that we pick one colloid at the center and sum over all the neighbors. The colloid-colloid  interaction potential is - at least in part - connected with  electrostatic interactions \cite{Kung} and thus follows from the electrostatic free energy with colloidal charges as external sources \cite{Dobnikar_2003}.

We now cast the Hookeian elastic energy in a different form by inserting the symmetrized deformation tensor $u_{ik} = \frac12 (\partial_i u_k + \partial_k u_i)$ into Eq. (\ref{equela0}), while using the Gauss theorem for the terms that can be cast into the form of a divergence. We thus obtain the elastic free energy as a sum of a volume integral and a surface integral, where the former contains  only the scalar invariants composed of $\bnabla\cdot {\bm u}$ and $\bnabla \times \bm u$, plus some irrelevant surface terms \footnote{$ \mu \oint_{S} {\bm n} \cdot \left[ {\textstyle\frac12} \bnabla {\bm u}^2- {\bm u} \times (\bnabla \times {\bm u}) - {\bm u} (\bnabla\cdot{\bm u})\right] {\rm d}S $}. This is of course possible only in the case of an isotropic elastic medium. Since the colloid crystal is an anisotropic elastic medium, this is an approximation which ignores the directional aspects of the elastically mediated interaction, but allows for a simplified derivation of the salient features of the positional dependence. 

In addition, the source term of Eq. (\ref{equela0}), corresponding to a point-like interstitial test particle in the elastic medium, can be modified taking into account that the point-like  body force can be written as ${\bm f} = {v_0 (\lambda + 2\mu) \bnabla \rho_e}$ \cite{Teodosiu1982,lechner2009point}, where $v_0$ is the local volume change induced by the point-like test particle, acting as a local dilation / contraction center; see Fig. \ref{Fig0}. { In fact,} $v_0$ of a point-like interstitial test particle plays the same role in elastic interactions as $e_0$ of a point-like charge plays in the electrostatic interactions, and the relaxation of the deformation away from the test particle is described by the Navier equation, and of the electrostatic potential by the PB equation.  

Furthermore it is clear that for an infinite sample volume one can derive the equality 
\begin{eqnarray}
\int_V {\rm d}^3{\bm x}~\bnabla\rho_e \cdot {\bm u} \longrightarrow  - \int_V {\rm d}^3{\bm x}~\rho_e \bnabla\cdot{\bm u},
\end{eqnarray}
which allows us to write the elastic source term in an analogous form as in the electrostatic free energy Eq. (\ref{FES1}). 
Effectively the same form of the elastic coupling,  based on a 2D point interstitial defect modelled by two orthogonal pairs of forces, has been used in Ref. \cite{lechner2009point}. 
The couplings of the elastic displacement field and the electrostatic potential to the density of the interstitial test particles $\rho_e$  are thus completely analogous and given by
\begin{eqnarray}
v_0 (\lambda + 2\mu) \int_V {\rm d}^3{\bm x}~  \rho_e  \bnabla\cdot{\bm u} \quad {\rm and} \quad 
 e_0 \int_V {\rm d}^3{\bm x}~\rho_e \phi,
 \label{coupling1}
\end{eqnarray}
for the elastic and  the electrostatic case, respectively. Clearly $v_0 (\lambda + 2\mu)$ plays the role of an effective elastic "charge". 

With these {\sl provisos} the elastic energy of an isotropic elastic medium then assumes a much simplified final form 
\begin{eqnarray}
{\cal F}_{EL}[{\bm u}] = \int_V {\rm d}^3{\bm x} ~\Big[ {\textstyle \frac 12} (\lambda + 2 \mu) (\bnabla \cdot \bm u)^2 
+ {\textstyle \frac 12}  \mu  (\bnabla \times \bm u)^2  \nonumber\\
+ v (\lambda + 2\mu) \rho_e  \bnabla\cdot{\bm u}\Big].~~~~~~~~~~~~~~~
\label{equela}
\end{eqnarray}
Note the analogy, but also the fundamental difference between the electrostatic and the elastic free energies:  while the former one is formulated in terms of the scalar field and after an expansion to the second order exhibits (Debye) screening, the latter  is formulated in terms of a vector field and does not exhibit any screening, remaining "critical" (i.e. having an inifinite screening length) for all values of the parameters. This is a general property of the elastic Hamiltonians in soft condensed matter that furthermore implies Casimir-like phenomena in e.g. nematic, smectic and cholesteric liquid crystals \cite{RevModPhys.71.1233,Karimi2019}. 


To the electrostatic and elastic free energies we now add an explicit minimal coupling term, corresponding to an energy cost associated with dilatation/contraction of the colloidal lattice due to a local excess of positive or negative mobile charge. Since in an ideal crystal there is a one-to-one mapping of particles to lattice positions  \cite{PhysRevB.81.134110}, the relation between the density change, $\delta c$, of the colloidal lattice with mean density $c$, and the divergence of the displacement field can be written as \cite{Landau}
\begin{eqnarray}
\delta c = - c ~\bnabla\!\cdot\!{\bm u}.
\end{eqnarray}
In our model we assume that the local density change of the colloidal lattice is a consequence of the local mobile charge density mismatch proportional to $(n_+ - n_-)$, and therefore the  { electrostatic-elastic} minimal coupling term can be assumed to have the form
\begin{eqnarray}
{\cal F}_{C}[ n_{\pm}, {\bm u}] = v_0 ~(\lambda + 2\mu) \int_V {\rm d}^3{\bm x} ~(n_+ - n_-) ~\bnabla\!\cdot\!{\bm u},
\end{eqnarray}
consistent with Eq. \ref{coupling1} and based upon assumption that the density change for positive and negative ions are the same but of opposite signs.   
We additionally assumed that the elastic coupling is the same - except for the sign - for both types of mobile charges, so there is only a single  coupling constant. The above coupling term is just the first term in a series of { higher-order symmetry-permitted} scalar invariants that we will not consider explicitly in this work. 

Gathering all the  contributions to the free energy in this minimal model of  { electrostatic-elastic} coupling, the total free energy is then given by 
\begin{eqnarray}
{\cal F} &=& {\cal F}_{ES}[n_{\pm}, \phi] + {\cal F}_{EL}[{\bm u}] + {\cal F}_{C}[n_{\pm}, {\bm u}] \nonumber\\
&=& \int_V {\rm d}^3{\bm x}~g(n_{\pm}, \phi, {\bm u}),
\label{freetot}
\end{eqnarray}
and the thermodynamic equilibrium is described by the corresponding Euler-Lagrange (EL) variational equations that are of the form
\begin{eqnarray}
\frac{\partial g}{\partial n_{\pm}} = 0 
\quad {\rm and} \quad
\frac{\partial g}{\partial \phi} - \bnabla\cdot\left(\frac{\partial g}{\partial \bnabla\phi}\right) = 0, \end{eqnarray}
as well as
\begin{eqnarray}
\frac{\partial g}{\partial u_i}  - \nabla_k\left(\frac{\partial g}{\partial ~\nabla_k {u_i}}\right) = 0.
\end{eqnarray}
The explicit form of these equations is as follows. For the mobile charges 
\begin{eqnarray}
&& n_+: \quad e_0 \phi  + {v_0 (\lambda + 2 \mu)} \bnabla\!\cdot\!{\bm u} + k_{\rm B}T \ln{n_+a^3} - \mu_0 = 0, \nonumber\\
&& n_-: \quad - e_0 \phi - {v_0 (\lambda + 2 \mu)} \bnabla\!\cdot\!{\bm u} + k_{\rm B}T \ln{n_-a^3} - \mu_0 = 0.~~~~~~
\end{eqnarray}
For the electrostatic potential we derive a modified Poisson equation and for the elastic displacement we derive a modified Navier equation 
\begin{eqnarray}
\phi &:& e_0 \rho_e + e_0(n_+-n_-) - \bnabla\!\cdot\Big( - \varepsilon\bnabla \phi  \Big) = 0,  \nonumber\\
{\bm u} &:& v_0 (\lambda + 2 \mu) \left( \bnabla \rho_e  +  \bnabla ( n_+-n_-)\right) + \mu \nabla^2 \bm u  \nonumber\\
&& ~~~~~~~~~~~~~~ +  (\lambda + \mu) \bnabla (\bnabla \cdot \bm u)= 0. 
\label{ELEa}
\end{eqnarray}
These equations can be cast into a much simpler form by introducing the charge density of the mobile ions 
\begin{eqnarray}
e_0 (n_+-n_-) &\equiv& \rho(\phi, {\bm u}) = \nonumber\\
& -&\frac{2 e_0 ~{\rm e}^{\beta \mu_0}}{a^3} \sinh\beta \left[e_0 \phi + v_0 (\lambda + 2 \mu)  \bnabla\!\cdot\!{\bm u} \right], ~~~~~~
\label{density1}
\end{eqnarray}
Note the difference: $\rho$ is the mean-field charge density while $\rho_e$ is the external test particle source density. 

The first equation in Eq~(\ref{ELEa})  presents a generalization of the PB equation, while the second one generalizes the Navier equation. Without the source term, the mean-field solution corresponds to the bulk state of vanishing electrostatic potential and vanishing elastic displacement.

We now invoke the standard Helmholtz decomposition \cite{lechner2009point} for the elastic displacement field {in terms of the scalar ($\Phi$) and vector ($\bm A$) elastic potentials }   
\begin{eqnarray}
{\bm u} = \bnabla \Phi + \bnabla \times {\bm A},
\label{Helmholtz}
\end{eqnarray}
representing the curl free longitudinal and the divergence free transversal part of the elastic deformation. Since all the {   { electrostatic-elastic}} couplings in the free energy Eq. \ref{freetot} are in terms of the $\bnabla\cdot {\bm u}$, we can assume the trivial solution ${\bm A} = const.$ \cite{lechner2009point} for an infinite sample. In addition, in the spirit of the {  DH} approximation \cite{Mar21},  we linearize the charge density to obtain
\begin{eqnarray}
\rho(\phi, {\bm u}) \simeq - 2 e_0 \lambda_0{\Big(e_0 \phi + v_0 (\lambda + 2 \mu)  \bnabla \cdot {\bm u}\Big)},
\label{bgak2}
\end{eqnarray}
with $\lambda_0 = \frac{\beta {\rm e}^{\beta \mu_0}}{a^3} = \beta n_0$, where $n_0$ is the density in the bulk reservoir. Just as in the case of the standard DH theory of Coulomb fluids, the  linearization {\sl Ansatz} is consistent if 
\begin{eqnarray}
\beta e_0 \phi \ll 1  \qquad {\rm and} \qquad \beta v_0 (\lambda + 2 \mu)  \bnabla\!\cdot\!{\bm u} \ll 1. 
\label{linear}
\end{eqnarray}
Since for point-like test particles both the electrostatic and the elastic { potentials} formally diverge at the source, see next section for details, one would need to introduce a finite size core inside which the theory breaks down, which actually  means one should use effective values for both $e_0$ and $v_0$.     

This linearization furthermore implies that the electrostatic free energy Eq~(\ref{FES1}) is expanded to the second order in the charge density and/or potential, effectively leading to Debye screening of electrostatics. Since,  as we already noted, the elastic energy Eq~(\ref{equela}) is "critical" and implies no separate elastic screening, one can expect that the coupled description will also yield only one type of electrostatic screening. 

Taking into account the linearization and the Helmholtz decomposition of the elastic displacement we end up with the following coupled set of linear equations
\begin{eqnarray}
&&e_0 \rho_e  = 2 \lambda_0 {e_0}^2 \phi - \varepsilon \nabla^2 \phi  +  2 \lambda_0 e_0 v_0 (\lambda + 2 \mu)  ~\nabla^2 \Phi, ~~~~~ \nonumber\\
&&\quad {  v_0 \rho_e =  2 \lambda_0 v_0  e_0 \phi + \left[ 2 \lambda_0 v_0^2 (\lambda+2\mu)-1\right] \nabla^2\Phi},
\label{bgak3}
\end{eqnarray}
where the {   { electrostatic-elastic}} coupling terms are characterized by the product $e_0 v_0$. The solution of these two equations gives the electrostatic potential and the elastic potential in the system up to the lowest order {   { electrostatic-elastic}} coupling terms.

\section{Electrostatic  and elastic deformation fields}
\label{sec:profile}

The EL equations Eqs. \ref{bgak3} can now be solved explicitly for a point-like test particle with an electrostatic charge $e_0$, and an elastic "charge" $v_0 (\lambda + 2 \mu)$, described by a delta function-like density $\rho_e = \delta({\bm r})$ located at the origin. The solution yields the spatial relaxation of the mobile ion charge density, $\bnabla \cdot {\bm E} = - \varepsilon \nabla^2 \phi$, and the colloid lattice relative density change , $\delta c/c = \bnabla \cdot {\bm u}$, away from the origin.

The solution of this set of linear equations can be obtained straightforwardly and we omit the unnecessary details. The electrostatic potential, proportional to the electrostatic charge of the test particle $e_0$, has the form
\begin{eqnarray}
\phi(r) = \frac{e_0~\alpha}{4 \pi \varepsilon} \frac{{\rm e}^{-\kappa r}}{r},
\label{phi}
\end{eqnarray}
with 
\begin{eqnarray}
\alpha = \frac{1}{1 - \xi} \quad {\rm and} \quad \kappa^2 = \alpha \kappa_0^2,
\end{eqnarray}
{with $\xi = 2 \beta n_0 v_0^2 (\lambda + 2\mu)$ the  { electrostatic-elastic}  coupling constant  and}  $\kappa_0^2 = {2 \lambda_0 e_0^2}/{\varepsilon}$ the inverse square of the standard Debye screening length. Since $\xi \geq 0$ it follows that the screening length with  { electrostatic-elastic} coupling decreases compared to the standard Debye screening length. 
The electrostatic field is then  obtained from 
\begin{equation}
{\bm E} = - \bnabla \phi = \frac{e_0~\alpha}{4 \pi \varepsilon} \left( 1 + \kappa r\right)\frac{{\rm e}^{-\kappa r}}{r^2} \frac{\bm r}{r},
\end{equation}
and describes the  spatial relaxation of the mobile ion charge density proportional to $\bnabla \cdot {\bm E}$.  
The elastic deformation potential, proportional to the elastic "charge" $v_0 (\lambda + 2\mu)$,  has the form
\begin{eqnarray}
{  \Phi(r) = \frac{v_0 \alpha}{4\pi} \frac{{\rm e}^{-\kappa r}}{r}},
\end{eqnarray}
and taking into account ${\bm u} = \bnabla \Phi$, we { have} the elastic deformation vector in the medium as
\begin{eqnarray}
{\bm u}\cdot \frac{\bm r}{r} = - \frac{v_0 \alpha (1+\kappa r)}{4\pi} \frac{{\rm e}^{-\kappa r}}{r^2},
\end{eqnarray}
describing the spatial relaxation of the colloid lattice relative density
change, $\bnabla \cdot {\bm u}$, as
\begin{eqnarray}
\bnabla\cdot{\bm u} = \nabla^2 \Phi = {v_0 \alpha} \frac{\kappa^2 {\rm e}^{-\kappa r}}{4\pi r}. \label{divu1}
\end{eqnarray}
Recalling the discussion on the limits of linearization in the previous section, we can now write the solutions for the point-like test particles as
\begin{eqnarray}
 &&   \beta e_0 \phi(r) = 2 (\kappa \ell_{\rm B}) \alpha \frac{{\rm e}^{-\kappa r}}{\kappa r}, \nonumber\\
&& \beta v_0 (\lambda + 2\mu) \bnabla\cdot {\bm u}(r) = 2 (\kappa \ell_{\rm E})^3 \alpha \frac{{\rm e}^{-\kappa r}}{\kappa r},
\label{lims}
\end{eqnarray}
where we introduced two characteristic length scales, the electrostatic Bjerrum length
\begin{eqnarray}
{  \ell_{\rm B} = {\beta e_0^2 }/(8\pi \varepsilon),}
\end{eqnarray}
and an elastic characteristic length given by
\begin{eqnarray}
{  \ell_{\rm E}^3 = \beta v_0^2 (\lambda + 2\mu)/(8\pi).}
\end{eqnarray}
Note the  { electrostatic-elastic} analogy in the definition of both lengthscales. We get an estimate for both lengths by assuming an aqueous solvent {  $\varepsilon =\varepsilon_{\rm r}\varepsilon_0 \sim 78\varepsilon_0$} with { $\ell_{\rm B} \sim 0.7$ nm, and a colloidal crystal bulk modulus $(\lambda + 2 \mu)\sim 100$ {\textrm dyne}$/${\textrm cm}$^2$} implying that $\ell_{\rm E} \sim ( 10^{-5} v_0^2)^{1/3}$ if $v_0$ is in nm. Consequently $\ell_{\rm E}$ and $\ell_{\rm B}$ would be comparable for $v_0$ being a fraction of a colloid lattice unit cell of $\sim$ $\mu$m lattice spacing.  In this case the validity of the linearization {\sl Ansatz} can then be expressed as $(\kappa \ell_{\rm B}) \alpha ({\rm e}^{-\kappa a})/(\kappa a)< 1$, which implies also $(\kappa \ell_{\rm E})^3 \alpha ({\rm e}^{-\kappa a})/(\kappa a)< 1$, where $a$ would correspond to an effective core radius of the interstitial test particle.

We next calculate the total interaction energy between two otherwise identical point-like test  sources with the same electrostatic charge and elastic "charge", 
one located at ${{\bm x}}_1$ and the other one at ${{\bm x}}_2$, with $\vert {{\bm x}}_1 - {{\bm x}}_2\vert = D$. From the total free energy and taking into account the linearized mean-field equations Eq~(\ref{bgak3}), we obtain
\begin{eqnarray}
{\cal F} &=& {\textstyle\frac{1}{2}}\int_V {\rm d}^3{\bm x} ~\Big( e_0 \rho_e \phi  +  v_0 (\lambda + 2\mu) \rho_e \bnabla\!\cdot\!{\bm u}\Big),~~~ 
\label{sphenergy}
\end{eqnarray}
which can be evaluated explicitly. Note that $\rho_e, \phi$ and $\bm u$ refer to the sum of all the source(s) and all the field(s) and thus necessarily contain also the self energies, however, we will be interested only in the part that depends on the separation $D$ between the sources. Using the second equation in Eq~(\ref{bgak3}) we derive Eq~(\ref{sphenergy}) in the final form
\begin{eqnarray}
{\cal F}(D) &=& \frac{{e_0}^2 \alpha^2 }{{  8} \pi \varepsilon} \frac{{\rm e}^{-\kappa D}}{D} = \frac{ {e_0}^2 \alpha^2}{{  8} \pi \varepsilon} \frac{{\rm e}^{-\sqrt{\alpha} ~\kappa_0 D}}{D}.
\end{eqnarray}
For non-zero $\xi$, it follows from its definition that $\alpha \geq 1$, while the decay length is obviously smaller (shorter ranged) than the bare Debye length, $\kappa \geq \kappa_0$. This follows from the fact (see Sec. 4) that the inverse square of the screening length is proportional to the charge density response function. Consequently, any terms additive in free energy, like the elastic terms in our case, add to the  charge density response function thus decreasing the effective screening length.

The interaction energy ${\cal F}(D)$ can be either larger or smaller then the standard (no background colloidal lattice) DH screened electrostatic interaction
\begin{eqnarray}
{\cal F}_{\rm DH} (D) = \frac{{e_0}^2}{{  8} \pi \varepsilon}~\frac{{\rm e}^{-\kappa_0 D}}{D},
\label{DBenergy}
\end{eqnarray}
depending on the interplay of the increased magnitude and increased screening. The {   { electrostatic-elastic}} coupling thus creates a non-trivial modification of the interactions between interstitial inclusions.

\section{Fluctuations around mean-field solution}

Notably, the solutions of the linearized mean-field equations, as derived above, can in principle be extended to a negative value of the coupling constant $\alpha$, or equivalently to imaginary screening length leading to oscillatory field solutions. Would this be real? An important issue that needs to be addressed at this point is therefore the range of validity of the mean-field solution that we derived above. The standard approach is to investigate the fluctuations around this solution, governed by the field Hessian, and ascertain that all the eigenvalues of the Hessian are positive.

In order to assess the role of fluctuations around the mean-field (MF) solution, $\delta {\bm u}$ and $\delta \phi$, we need to analyze the deviations from the mean-field solution quantified by ${\bm u} = {\bm u}_{\rm MF} + \delta {\bm u}$ and $\phi = \phi_{\rm MF} + \delta \phi$, where ${\bm u}_{\rm MF}, \phi_{\rm MF}$ are the solutions of the EL equations derived above. In order to proceed, we first note that the mean-field free energy Eq~(\ref{freetot})  can be cast into the form
\begin{eqnarray}
{\cal F}[\phi, {\bm u}] &=& \int_V {\rm d}^3{\bm x}~ \Big( -{\textstyle\frac12} \varepsilon \left(\bnabla \phi\right)^2 +{\textstyle \frac 12} (\lambda + 2 \mu) (\bnabla\cdot\bm u)^2 \nonumber \\
&& - 2 k_{\rm B}T n_0 \cosh{\beta \left(e_0 \phi + v_0 (\lambda + 2 \mu)  \bnabla\!\cdot\!{\bm u}\right)}\Big),~~~~~~
\label{fluc0}
\end{eqnarray}
which now contains only the field variables. {Without the elastic displacement terms this free energy would correspond to the well known alternative form of the PB free energy that contains only the mean-field electrostatic potential \cite{Verwey}}.

The exact partition function now has to be written as a functional integral over the imaginary field $i \phi$ \cite{WIEGEL197557, Blossey_2018} and the field $\bm u$. Since we are interested only in the limit of stability of the mean-field solution, we expand this field  partition function to the second order in $\delta \phi$ and $\delta {\bm u}$ around the saddle point of the field action, corresponding to the evaluation of the field Hessian, which will  fortunately turn out to be rather simple to analyze,  and will consequently answer our original question on the range of validity of the mean-field theory.

Expanding the total field free energy to the second order we are then led to a deviation from the mean-field value of the free energy for the electrostatic potential and the longitudinal component of the deformation field
\begin{widetext}
\begin{eqnarray}
&& ~~~~~~~~~~~~~{\cal F}[\phi_{\rm MF} + \delta \phi, {\bm u}_{\rm MF} + \delta {\bm u}] - {\cal F}[ \phi_{\rm MF}, {\bm u}_{\rm MF}] \nonumber\\ && = 
 \int_V {\rm d}^3{\bm x}~ \Big( -{\textstyle\frac12} \varepsilon \left(\bnabla \delta\phi\right)^2 + \frac{\partial \rho}{\partial \phi} \bigg\rvert_{\rm MF} \!\!\!\!\delta\phi^2  + ~{\textstyle \frac 12} (\lambda + 2 \mu) (\bnabla \cdot \delta\bm u)^2  + {\textstyle \frac 12} v_0 (\lambda + 2 \mu) \frac{\partial \rho}{\partial  (\bnabla\!\cdot\!\delta\bm u)} \bigg\rvert_{\rm MF} \!\!\!\!(\bnabla\!\cdot\!\delta\bm u)^2 \\ && \quad + ~ {\rm i} ~v_0 (\lambda + 2 \mu) \frac{\partial \rho}{\partial \phi} \bigg\rvert_{\rm MF}\!\!\!\!\delta\phi ~(\bnabla\!\cdot\!\delta\bm u) \Big), \nonumber
\label{fluc1}
\end{eqnarray}
\end{widetext}
since the transverse component of $\delta{\bm u}$ is not coupled to electrostatics. Above we introduced the derivatives of the mean-field charge density Eq~(\ref{density1}). As already noted, the $i$ in front of the term linear in $\delta \phi$ is stemming from the functional integral representation of the Coulomb fluid partition function  \cite{WIEGEL197557,Blossey_2018}. In addition, because of Eq~(\ref{density1}), we have
\begin{eqnarray}
e_0 \frac{\partial \rho}{\partial  (\bnabla\!\cdot\!\delta\bm u)} = v_0 (\lambda + 2 \mu) \frac{\partial \rho}{\partial \phi}.
\end{eqnarray}
The derivative of the charge density, $\frac{\partial \rho(\phi)}{\partial \phi}$, by definition the capacitance density response function, is in fact proportional to the square of the inverse Debye screening length \cite{AVNI201970}
\begin{eqnarray}
 \kappa_0^2 = - \frac{1}{\varepsilon} \frac{\partial \rho(\phi)}{\partial \phi} \bigg\rvert_{\rm MF} = \frac{2 \lambda_0 e_0^2}{\varepsilon}.
 \end{eqnarray}
The longitudinal component of the elastic displacement vector derivatives yields a Gaussian integral and can be integrated out from the partition function, yielding a purely electrostatic { expression} with a renormalized screening length
\begin{eqnarray}
&& \kappa^2 \longrightarrow 
{\kappa_0^2} \alpha = \kappa_0^2 \frac{1}{1-\xi}.
\end{eqnarray}
This of course tallies with the screening length derived from the mean-field equations.

\begin{figure*}[t!]
\includegraphics[width=16cm]{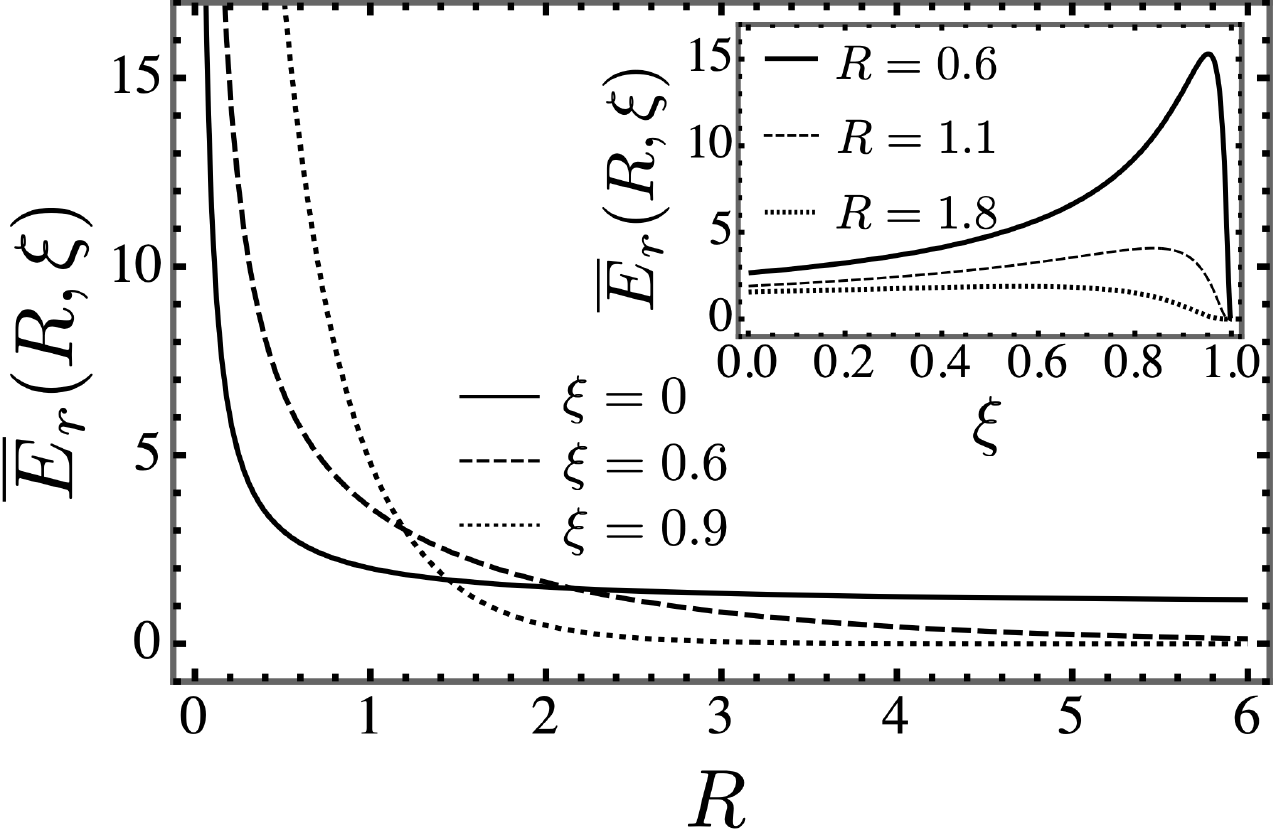}
\caption{The behaviors of the dimensionless electrostatic field ${\overline E}(R, \xi)$, Eq~(\ref{EquE}), as a function of both relevant variables $R = \kappa_0 r$ and $\xi = 2 \beta n_0 v_0^2 (\lambda + 2\mu)$ {  (inset)}. Clearly, the dependence on the coupling parameter $\xi$ is non-monotonic in the inset, but the dependence on the dimensionless radial distance from the point source $R$ is monotonic, and { the decay} is the fastest for the pure {  electrostatic} system $\xi = 0$.} 
\label{ebar}
\end{figure*}

Since the condition that the eigenvalues of the Hessian be positive is thus reduced to the condition that the screening length is real, the stability of the mean-field solution is defined by $1-\xi \geq 0$, with strict identity corresponding to the limit of stability. 

Let us investigate then the consequences of this result. {From the definition of the elastic constants, Eq. (\ref{consts}), it follows that $\xi \sim (n_0 a^3) (v_0/a^3)^2 (\beta \upsilon''(a) a^2)$, where $a$ is the colloidal lattice spacing and the colloid-colloid interaction  potential was assumed to be short ranged. The strength of this interaction  potential depends on the colloidal charge, which however cannot be measured directly \cite{Kung}. Taking instead the measured colloidal crystal bulk modulus $(\lambda + 2 \mu)\sim 100$ {\textrm dyne}$/${\textrm cm}$^2$, the regime of $\xi \leq 1$ holds if $\beta \upsilon''(a) \ll 1/a^2$ and $n_0 \ll 1/a^3$. If on the other hand these criteria are not satisfied and the colloidal crystal is strongly charged, the mean-filed would cease to be a valid description and following the general analogy with Coulomb fluids \cite{naji2011exotic} one would need to formulate a strong coupling theory for this problem.}  

Based on the above analysis, the applicability of the mean field theory is therefore limited by the requirement that $\xi \leq 1$, which 
can be reduced to 
\begin{equation}
\xi = (n_0a^3) \left( \frac{v_0}{a^3}\right)^2 (\kappa a)^2 \beta v(a) \ll 1,
\end{equation}
while the validity of the linearization {\sl Ansatz}, Eq. \ref{linear}, together with the solution of the linearized equations, Eqs. \ref{divu1} and \ref{phi}, can be expressed as 
\begin{equation}
   \xi ~\alpha \left( \frac{a}{b}\right) {\rm e}^{-\kappa b}\ll 1, 
\end{equation}    
where $b$ would correspond to an effective core radius of the interstitial test particle. Clearly assuming $\xi \ll 1$, so that mean field is valid, implies that the linearization of the mean field equations is also allowed if the screening length is much larger then the core radius of the colloid. Assuming the values relevant to experiment \cite{Liu2022}, $b \sim .1 \mu m$, $a \sim 1-4 \mu m$,  and the effective salt concentration estimated from the dissociation of the colloidal particles to be below $\sim \mu M$ both above conditions are satisfied.

As we noted above, outside the stable regime of parameter values the MF solutions would formally acquire a periodic component, perhaps indicating a delocalized-to-localized transition in the mobile ion density. Remains to be explored either within a strong-coupling framework or detailed simulations that fully include the electrostatic interactions.

\section{Results and discussion}

We now analyze the general results derived above for the electrostatic field and elastic displacement field of a single test particle, and also the interaction between two test particles  located at a finite separation, in terms of the parameters of the model. The point-like test particles are characterized by an electrostatic charge $e_0$ and an elastic "{  charge}" $v_0 (\lambda + 2 \mu)$ that are both taken as parameters of the model.

\begin{figure*}[t!]
\includegraphics[width=16cm]{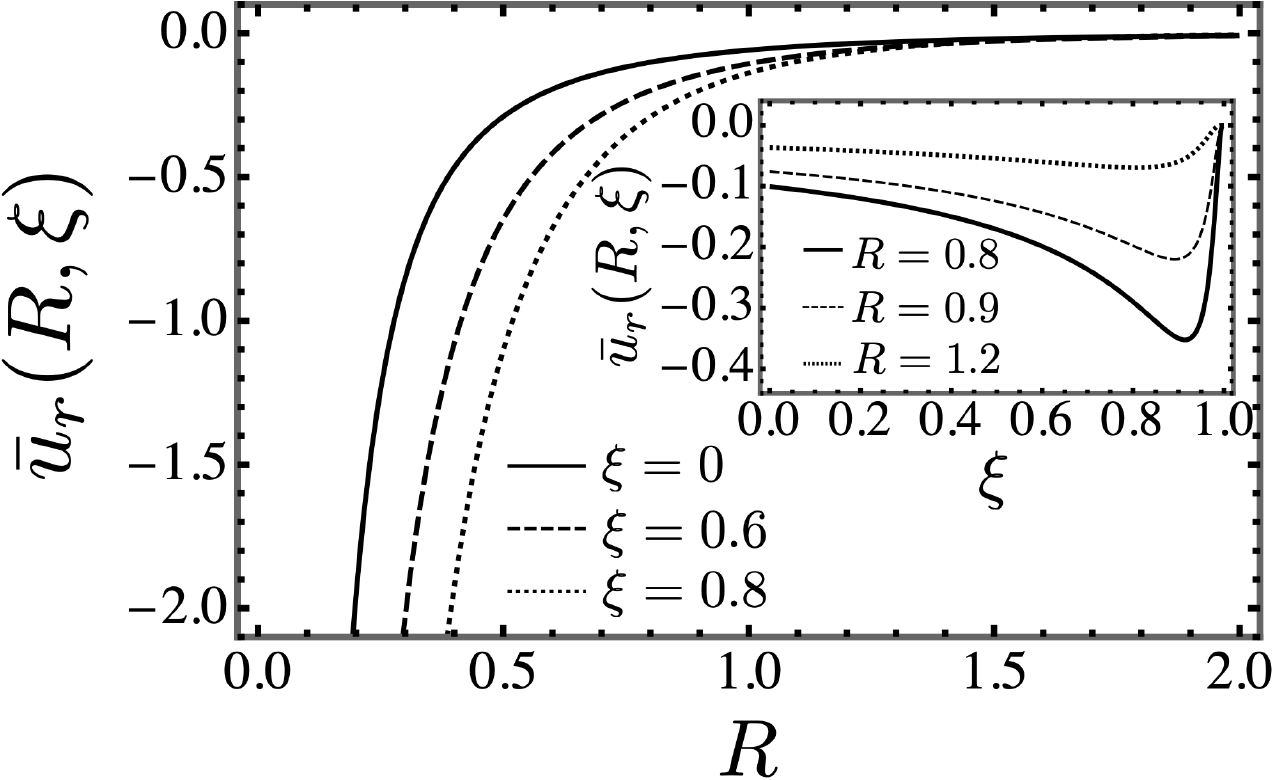}
\caption{The behaviors of the dimensionless radial elastic displacement field ${\bar{u}_r}(R, \xi)$, Eq~(\ref{EquUr}), as a function of both variables $R = \kappa_0 r$ and $\xi = 2 \beta n_0  v_0^2 (\lambda + 2\mu)$ {  (inset)}. Clearly, the dependence on the coupling parameter $\xi$ is non-monotonic, while the dependence on the dimensionless radial distance from the point source is monotonic, and { the increase} is the fastest for the pure electrostatic system.}
\label{ur}
\end{figure*}

The radial component of the electrostatic field of a single point source can be cast into the form
\begin{eqnarray}
{E}_r(R , \xi) &=&  
-\frac{{\rm d} (\beta e_0\phi(r))}{{\rm d} (\kappa_0 r)} \nonumber\\
&=& 
\frac{{e_0}^2 ~{\rm e}^{-\kappa_0 r}}{4 \pi \varepsilon~ r}\frac{1}{1-\xi}\left(\frac{1+\frac{\kappa_0 r}{\sqrt{1-\xi}}}{\kappa_0 r}\right){\rm e}^{-\left(\frac{1}{\sqrt{1-\xi}}-1\right)\kappa_0 r} \nonumber\\
&=& {\cal F}_{\rm DH}(r)\overline{E}_r(R, \xi).
\label{EquE}
\end{eqnarray}
where the radial electrostatic field $E_r(R, \xi)$ can be written in a dimensionless form ${\overline{E}_r(R, \xi)}$ as a function of variables $R = \kappa_0 r$ and $\xi = 2 \beta n_0 v_0^2 (\lambda + 2\mu)$. Its dependence $E_r(R, \xi)$ is presented in Fig. \ref{ebar}.

Next we characterize the radial component of the elastic displacement of the colloidal lattice as follows
\begin{eqnarray}
u_r(R, \xi) &=& {\bm u}(r) \cdot \frac{{\bm r}}{r}\nonumber\\
&=& -\frac{v_0}{4\pi  r^2}\left(\frac{1}{1-\xi}\right)\left(1+\frac{\kappa_0 r}{\sqrt{1-\xi}}\right){\rm e}^{-\frac{\kappa_0 r}{\sqrt{1-\xi}}}\nonumber\\
&=& v_0 \kappa_0^2 \bar{u}_r (R,\xi)
\label{EquUr}
\end{eqnarray}
where ${\bar{u}_r}(R, \xi)$ is a dimensionless form as a function of variables $R = \kappa_0 r$ and $\xi = 2 \beta n_0 v_0^2 (\lambda + 2\mu)$ and quantifies the elastic displacement of the colloidal lattice induced by the electrostatic and elastic interactions. Its dependence ${\bar{u}_r}(R, \xi)$ is presented in Fig. \ref{ur}.

We finally analyze the consequences of the  { electrostatic-elastic} coupling on the interaction free energy, Eq. (\ref{sphenergy}), between two test particles  characterized by the same $e_0, v_0$ but located $D$ apart. Comparing the lattice-mediated electrostatic interaction with the standard DH interaction, we write the result in the following form 
\begin{eqnarray}
\Delta {\cal F}(D) &\equiv&  {\cal F}(D) - {\cal F}_{\rm DH}(D) \nonumber\\ 
&=& \frac{{e_0}^2 {\rm e}^{-\kappa_0 D}}{{  8} \pi \varepsilon D}  \left[\left(\frac{1}{1-\xi}\right)^2 {{\rm e}^{-(\kappa-\kappa_0) D}} - 1\right] \nonumber\\
&=& {\cal F}_{\rm DH}(D) \overline{\cal W}({\tilde R}, \xi).
\label{EquW}
\end{eqnarray}
where ${\tilde R} = \kappa_0 D$, and $\overline{\cal W}({\tilde R}, \xi)$, the relative change with respect to the DH interaction, can be cast into a dimensionless form as a function of dimensionless variables ${\tilde R}$ and $\xi$, quantifying the relative change in the DH interaction energy due to  { electrostatic-elastic} coupling. It can be either positive, ${\cal F}(D) \geq {\cal F}_{\rm DH}(D)$,  corresponding to repulsive elastic coupling induced interactions,  or negative, ${\cal F}(D) \leq {\cal F}_{\rm DH}(D)$, corresponding to attractive elastic coupling induced interactions, dependent on both the coupling strength $\xi$ and the distance $D$. 

\begin{figure*}[t!]
\includegraphics[width=16cm]{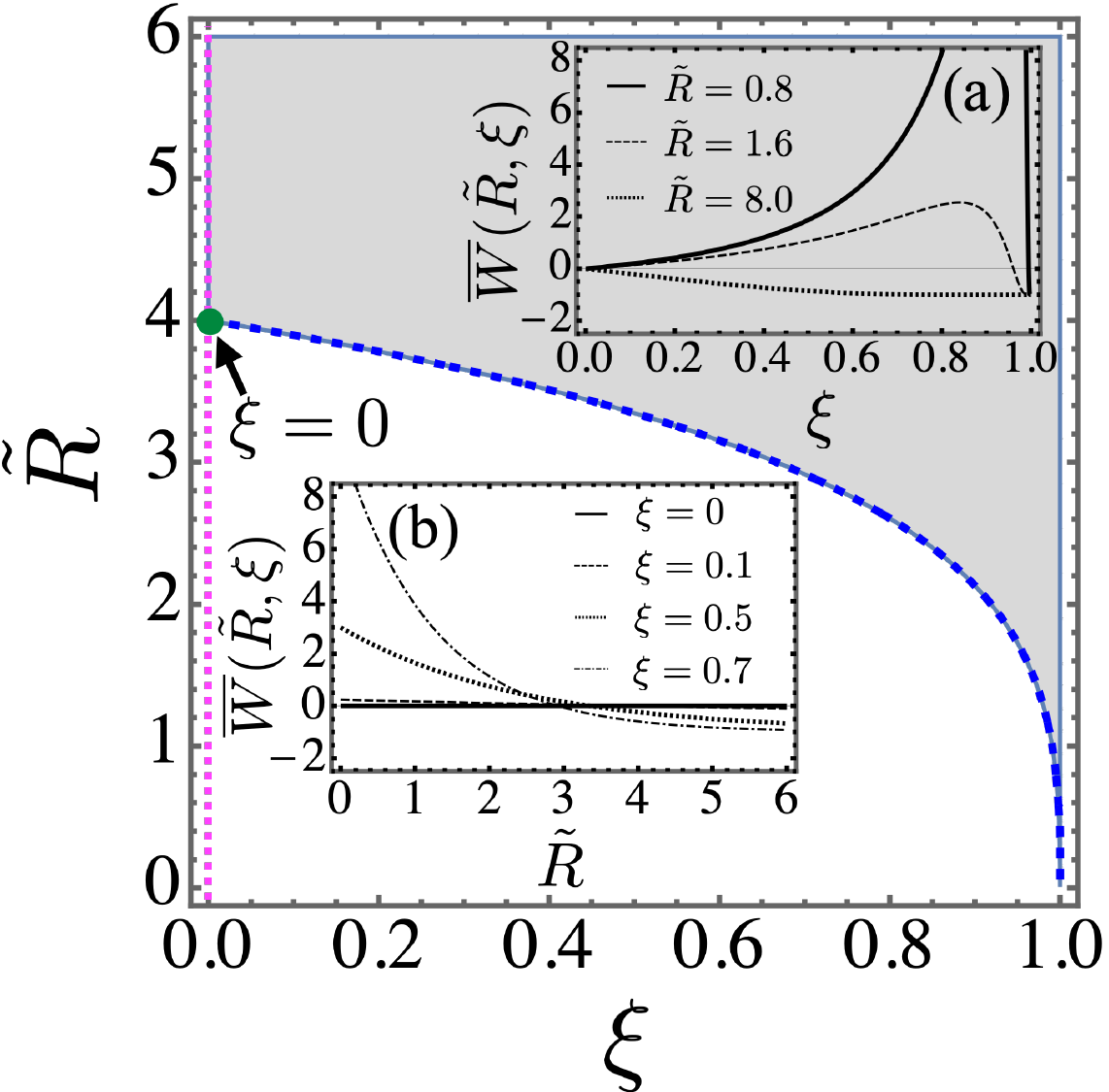}
\caption{Behaviors of the dimensionless total  { electrostatic-elastic} coupling free energy $\overline{\cal W}(\tilde{R}, \xi)$, Eq~(\ref{EquW}) as a function of both variables $\tilde{R} = \kappa_0 D$ and $\xi = 2 \beta n_0 v_0^2 (\lambda + 2\mu)$. Under a moderate dimensionless distance ${\tilde R}<4.0$, the dimensionless total energy shows a non-monotonic dependence on the  { electrostatic-elastic} coupling parameter $\xi$, but at large enough distance ${\tilde R}>4.0$, it changes to a monotonic decaying dependence on the coupling parameter $\xi$ and the interaction is purely attractive {  (Inset (a))}. The smallest attainable value of ${\cal W}(\tilde{R}, \xi)$ is obviously $-1$ which corresponds to the vanishing of the {   { electrostatic-elastic} coupling} interaction, with attractive contribution compensating the repulsive contribution. { But} the dependence on the dimensionless radial distance from the point source is decreasing, and { the decay} is {  slower as the  { electrostatic-elastic} coupling constant is weaker. Finally, it vanishes for the pure electrostatic} system {  $\xi=0$ (Inset (b))}. And the "phase diagram" in the parameter space delineates a repulsive region in an otherwise attractive sea.  {  The magenta dotted line shows our theory goes back to the standard DH theory for the coupling strength $\xi = 0$. The blue dashed line indicates a "critical line" $\tilde{R}^*$ as a function of $\xi$ separating the repulsive and attractive regions. A universal "critical" constant $\tilde{R}^*=4$ is indicated by a arrow for $\xi=0$.}
}
\label{figurestar}
\end{figure*}

{  In order to understand the phase diagram better, we take the logarithm of both sides of $\overline{\cal W}({\tilde R}, \xi)=0$ to further obtain
\begin{eqnarray}
-2\ln{\left(1-\xi\right)} = \left(\frac{1}{\sqrt{1-\xi}}-1\right){\tilde R}.
\label{EquW2}
\end{eqnarray}
Let $1/\sqrt{1-\xi}=1+\delta$, then the above equation can be rewritten as
\begin{eqnarray}
4\ln(1+\delta)=\delta{\tilde R}.
\label{EquW3}
\end{eqnarray}
With $\xi$ approaching $0$, $\delta$ also approaches $0$ and ${\tilde R}$ thus goes to a universal "critical" constant $\tilde{R}^*=4$, which does not depend on any parameters or variables (see the green dot pointed by an arrow in Fig. \ref{figurestar}.}

Inset (a) in Fig. \ref{figurestar} displays the dependence of  $\overline{\cal W}({\tilde R}, \xi)$ on $\xi$ for different values of ${\tilde R}$. Clearly $\overline{\cal W}({\tilde R}, \xi)$ corresponds to additional {   { electrostatic-elastic}} coupling generated interactions, apart from and on top of the usual DH interactions.  $\overline{\cal W}({\tilde R}, \xi) = -1$ indicates that there are no interactions, so the DH repulsion is completely compensated by the additional {   { electrostatic-elastic}} coupling interaction.  $\overline{\cal W}({\tilde R}, \xi) \geq 0$ implies additional interactions that are repulsive and $-1 \leq \overline{\cal W}({\tilde R}, \xi) \leq 0$ corresponds to additional attractive  {   { electrostatic-elastic}} coupling interactions that however do not overpower the underlying DH repulsion. $\overline{\cal W}({\tilde R}, \xi)$  starts off at zero at ${\xi} = 0$, and then depending on $\tilde R$ either develops a local negative minimum $\overline{\cal W} = -1$, for ${\tilde R}<4.0$, or spikes into a local positive maximum for ${\tilde R}>4.0$ until again settling down at $\overline{\cal W} = -1$ for $\xi = 1$. As can be furthermore seen from Fig. \ref{figurestar}, {  $\overline{\cal W}({\tilde R}, \xi) \geq 0$ is possible} only for ${\tilde R} \leq  {\tilde R}_c = 4.0$, which represents in some sense a {  " critical line" $\tilde{R}^*$ as a function of $\xi$} of the {   { electrostatic-elastic} coupling} interaction, separating a regime of ${\cal F}(D) \leq {\cal F}_{\rm DH}(D)$ from ${\cal F}(D) \geq {\cal F}_{\rm DH}(D)$. For ${\tilde R} > {\tilde R}_c$, {  $-1 \leq \overline{\cal W}({\tilde R}, \xi) \leq 0$ corresponds to} a regime where the {   { electrostatic-elastic}} coupling contributes a purely attractive additional interaction for all values of $\xi$.

Inset (b) in Fig. \ref{figurestar} displays a monotonic dependence of $\overline{\cal W}({\tilde R}, \xi)$ on ${\tilde R}$ for different values of $\xi$. Clearly, $\overline{\cal W}({\tilde R}, \xi)$ decays exponentially, but depending on $\xi$ it can show a regime where the {   { electrostatic-elastic}} coupling contributes a purely repulsive additional interaction for small ${\tilde R}$, reverting invariably to an attractive regime of additional interaction for large enough ${\tilde R}$. In addition, the larger $\xi$ is, the faster is the decay of the additional {   { electrostatic-elastic}} coupling generated interaction. The screening length can be in fact much smaller then the Debye length, so the {   { electrostatic-elastic}} coupling effect can be short range. 

A simple continuum model of  { electrostatic-elastic} coupling between mobile small  ions and a fixed colloidal lattice, based on the mean-field electrostatic energy and Hookeian elasticity, coupled by minimal coupling terms and solved in the linearized mean-field framework, can thus lead to an attractive additional  component in the overall interaction between two test particles in the colloidal lattice. While this lattice-mediated interaction  evaluated on the mean-field level  cannot overpower the underlying electrostatic  repulsion between charges, it can effectively eliminate it, {\sl i.e.} when $\overline{\cal W}({\tilde R}, \xi) = -1$. On the level of this theory the  { electrostatic-elastic} coupling can thus substantially diminish the repulsion between two mobile charges but cannot completely reverse its sign. 

{Our approach is necessarily approximate and model-dependent, as such being open to generalizations in different directions. While the continuum elastic lattice approximation allows for straightforward analytical  developments, the connection between the colloid-colloid interaction potential $\upsilon(\vert r_i - r_j\vert)$, see Eq. (\ref{collint}), and the interstitial Coulomb fluid is not direct, requiring an additional, even if very reasonable, assumption that it is of a screened Coulomb type. Note also that the colloid-colloid interaction potential depends on the assumed value of the dissociated charge of the colloids, that can include the effects of distorted double layers at high macroion volume fractions, but {  it} is also difficult to disentangle from experiments and even more from a detailed calculation based on explicit charge dissociation models \cite{AVNI201970}. Using phenomenological elastic constants as opposed to phenomenological effective colloidal charges seems in our view more appropriate and going beyond would necessitate a different formulation that would lose the intuitive clarity of Hookeian elasticity.}

There are two obvious generalizations that could be pursued in a more detailed description of this system. { The first would be to alter the model and bypass the continuum Hookeian lattice approximation. Instead one would consider explicit finite size colloids on a lattice, and solve the PB equation in the interstitial space with the proper boundary conditions \cite{Schmitz1999,Dobnikar_2003}, possibly including also charge regulation \cite{AVNI201970} at the surface of the macroions. While this could allow for a more detailed inclusion of the finite size effects \cite{Kornyshev2007}, the theory would have to be formulated differently possibly lacking a direct connection with Hookeian elasticity.} 

{The next step could be to go beyond the mean-field approximation.  As the colloids are strongly charged, the colloidal crystal can certainly be conceived as a {\sl strongly coupled electrostatic system} in the strict sense of the weak-strong coupling dichotomy  \cite{BOROUDJERDI2005129}. The strong coupling in this context refers to the coupling between mobile
ions and the lattice colloids, and the putative extension of our theory would then follow the very same pathway as in the standard strong coupling formalism \cite{Perspective}. It seems reasonable to expect that on the level of the strong coupling theory one would end up with a net attraction between test particles}, 
but with possibly even more complicated condensed structures of mobile charges within a fixed, charged colloidal lattice. 
Only after the electrostatics are properly included and modeled does the simulation make sense as a comparison with our theory. In addition, it would be interesting to compare not only the relative probability of tracer pairs in the background colloid lattice but also an actual interaction potential that could then be compared directly with Eq. \ref{EquW}.

Finally, our theory proposes a new mechanism for generating effective attractive interactions between equally charged ions in a colloid crystalline lattice that stems from the coupling between electrostatics and lattice elasticity, proposing a new mechanism that could prove to be important for colloid physics.

\begin{acknowledgments}
{  HW} is partially supported by the open research fund of Songshan Lake Materials Laboratory No. 2023SLABFN20, the General Program of National Natural Science Foundation of China (NSFC) under Grant No. 12374210 and the startup fund (Grant No. WIUCASQD2022005) from Wenzhou Institute University of Chinese Academy of Sciences (WIUCAS). {  Z-CO-Y} was supported by the Major Program of NSFC (Grant No. 22193032). {  RP} acknowledges the support of UCAS and funding from the Key Program of NSFC (Grant No.12034019).
\end{acknowledgments}




\bibliographystyle{apsrev4-2}
\bibliography{CTPreference}

\begin{thebibliography}{49}%
\makeatletter
\providecommand \@ifxundefined [1]{%
 \@ifx{#1\undefined}
}%
\providecommand \@ifnum [1]{%
 \ifnum #1\expandafter \@firstoftwo
 \else \expandafter \@secondoftwo
 \fi
}%
\providecommand \@ifx [1]{%
 \ifx #1\expandafter \@firstoftwo
 \else \expandafter \@secondoftwo
 \fi
}%
\providecommand \natexlab [1]{#1}%
\providecommand \enquote  [1]{``#1''}%
\providecommand \bibnamefont  [1]{#1}%
\providecommand \bibfnamefont [1]{#1}%
\providecommand \citenamefont [1]{#1}%
\providecommand \href@noop [0]{\@secondoftwo}%
\providecommand \href [0]{\begingroup \@sanitize@url \@href}%
\providecommand \@href[1]{\@@startlink{#1}\@@href}%
\providecommand \@@href[1]{\endgroup#1\@@endlink}%
\providecommand \@sanitize@url [0]{\catcode `\\12\catcode `\$12\catcode
  `\&12\catcode `\#12\catcode `\^12\catcode `\_12\catcode `\%12\relax}%
\providecommand \@@startlink[1]{}%
\providecommand \@@endlink[0]{}%
\providecommand \url  [0]{\begingroup\@sanitize@url \@url }%
\providecommand \@url [1]{\endgroup\@href {#1}{\urlprefix }}%
\providecommand \urlprefix  [0]{URL }%
\providecommand \Eprint [0]{\href }%
\providecommand \doibase [0]{https://doi.org/}%
\providecommand \selectlanguage [0]{\@gobble}%
\providecommand \bibinfo  [0]{\@secondoftwo}%
\providecommand \bibfield  [0]{\@secondoftwo}%
\providecommand \translation [1]{[#1]}%
\providecommand \BibitemOpen [0]{}%
\providecommand \bibitemStop [0]{}%
\providecommand \bibitemNoStop [0]{.\EOS\space}%
\providecommand \EOS [0]{\spacefactor3000\relax}%
\providecommand \BibitemShut  [1]{\csname bibitem#1\endcsname}%
\let\auto@bib@innerbib\@empty
\bibitem [{\citenamefont {Girard}\ \emph {et~al.}(2019)\citenamefont {Girard},
  \citenamefont {Wang}, \citenamefont {Du}, \citenamefont {Das}, \citenamefont
  {Huang}, \citenamefont {Dravid}, \citenamefont {Lee}, \citenamefont
  {Mirkin},\ and\ \citenamefont {de~la Cruz}}]{Olvera2019}%
  \BibitemOpen
  \bibfield  {author} {\bibinfo {author} {\bibfnamefont {M.}~\bibnamefont
  {Girard}}, \bibinfo {author} {\bibfnamefont {S.}~\bibnamefont {Wang}},
  \bibinfo {author} {\bibfnamefont {J.~S.}\ \bibnamefont {Du}}, \bibinfo
  {author} {\bibfnamefont {A.}~\bibnamefont {Das}}, \bibinfo {author}
  {\bibfnamefont {Z.}~\bibnamefont {Huang}}, \bibinfo {author} {\bibfnamefont
  {V.~P.}\ \bibnamefont {Dravid}}, \bibinfo {author} {\bibfnamefont
  {B.}~\bibnamefont {Lee}}, \bibinfo {author} {\bibfnamefont {C.~A.}\
  \bibnamefont {Mirkin}},\ and\ \bibinfo {author} {\bibfnamefont {M.~O.}\
  \bibnamefont {de~la Cruz}},\ }\href@noop {} {\bibfield  {journal} {\bibinfo
  {journal} {Science}\ }\textbf {\bibinfo {volume} {364}},\ \bibinfo {pages}
  {1174} (\bibinfo {year} {2019})}\BibitemShut {NoStop}%
\bibitem [{\citenamefont {Williams}\ and\ \citenamefont
  {Crandall}(1974)}]{WILLIAMS1974225}%
  \BibitemOpen
  \bibfield  {author} {\bibinfo {author} {\bibfnamefont {R.}~\bibnamefont
  {Williams}}\ and\ \bibinfo {author} {\bibfnamefont {R.}~\bibnamefont
  {Crandall}},\ }\href@noop {} {\bibfield  {journal} {\bibinfo  {journal}
  {Physics Letters A}\ }\textbf {\bibinfo {volume} {48}},\ \bibinfo {pages}
  {225} (\bibinfo {year} {1974})}\BibitemShut {NoStop}%
\bibitem [{\citenamefont {Leunissen}\ and\ \citenamefont {van
  Blaaderen}(2008)}]{Blaaderen}%
  \BibitemOpen
  \bibfield  {author} {\bibinfo {author} {\bibfnamefont {M.~E.}\ \bibnamefont
  {Leunissen}}\ and\ \bibinfo {author} {\bibfnamefont {A.}~\bibnamefont {van
  Blaaderen}},\ }\href@noop {} {\bibfield  {journal} {\bibinfo  {journal} {The
  Journal of Chemical Physics}\ }\textbf {\bibinfo {volume} {128}},\ \bibinfo
  {pages} {164509} (\bibinfo {year} {2008})}\BibitemShut {NoStop}%
\bibitem [{\citenamefont {Sirota}\ \emph {et~al.}(1989)\citenamefont {Sirota},
  \citenamefont {Ou-Yang}, \citenamefont {Sinha}, \citenamefont {Chaikin},
  \citenamefont {Axe},\ and\ \citenamefont {Fujii}}]{PhysRevLett.62.1524}%
  \BibitemOpen
  \bibfield  {author} {\bibinfo {author} {\bibfnamefont {E.~B.}\ \bibnamefont
  {Sirota}}, \bibinfo {author} {\bibfnamefont {H.~D.}\ \bibnamefont {Ou-Yang}},
  \bibinfo {author} {\bibfnamefont {S.~K.}\ \bibnamefont {Sinha}}, \bibinfo
  {author} {\bibfnamefont {P.~M.}\ \bibnamefont {Chaikin}}, \bibinfo {author}
  {\bibfnamefont {J.~D.}\ \bibnamefont {Axe}},\ and\ \bibinfo {author}
  {\bibfnamefont {Y.}~\bibnamefont {Fujii}},\ }\href
  {https://doi.org/10.1103/PhysRevLett.62.1524} {\bibfield  {journal} {\bibinfo
   {journal} {Phys. Rev. Lett.}\ }\textbf {\bibinfo {volume} {62}},\ \bibinfo
  {pages} {1524} (\bibinfo {year} {1989})}\BibitemShut {NoStop}%
\bibitem [{\citenamefont {Russell}\ \emph {et~al.}(2015)\citenamefont
  {Russell}, \citenamefont {Spaepen},\ and\ \citenamefont
  {Weitz}}]{PhysRevE.91.032310}%
  \BibitemOpen
  \bibfield  {author} {\bibinfo {author} {\bibfnamefont {E.~R.}\ \bibnamefont
  {Russell}}, \bibinfo {author} {\bibfnamefont {F.}~\bibnamefont {Spaepen}},\
  and\ \bibinfo {author} {\bibfnamefont {D.~A.}\ \bibnamefont {Weitz}},\ }\href
  {https://doi.org/10.1103/PhysRevE.91.032310} {\bibfield  {journal} {\bibinfo
  {journal} {Phys. Rev. E}\ }\textbf {\bibinfo {volume} {91}},\ \bibinfo
  {pages} {032310} (\bibinfo {year} {2015})}\BibitemShut {NoStop}%
\bibitem [{\citenamefont {Dotera}\ \emph {et~al.}(2014)\citenamefont {Dotera},
  \citenamefont {Oshiro},\ and\ \citenamefont {Ziherl}}]{Dotera2014}%
  \BibitemOpen
  \bibfield  {author} {\bibinfo {author} {\bibfnamefont {T.}~\bibnamefont
  {Dotera}}, \bibinfo {author} {\bibfnamefont {T.}~\bibnamefont {Oshiro}},\
  and\ \bibinfo {author} {\bibfnamefont {P.}~\bibnamefont {Ziherl}},\ }\href
  {https://doi.org/10.1038/nature12938} {\bibfield  {journal} {\bibinfo
  {journal} {Nature}\ }\textbf {\bibinfo {volume} {506}},\ \bibinfo {pages}
  {208} (\bibinfo {year} {2014})}\BibitemShut {NoStop}%
\bibitem [{\citenamefont {Groenewold}\ and\ \citenamefont
  {Kegel}(2001)}]{Kegel}%
  \BibitemOpen
  \bibfield  {author} {\bibinfo {author} {\bibfnamefont {J.}~\bibnamefont
  {Groenewold}}\ and\ \bibinfo {author} {\bibfnamefont {W.~K.}\ \bibnamefont
  {Kegel}},\ }\href {https://doi.org/10.1021/jp011646w} {\bibfield  {journal}
  {\bibinfo  {journal} {The Journal of Physical Chemistry B}\ }\textbf
  {\bibinfo {volume} {105}},\ \bibinfo {pages} {11702} (\bibinfo {year}
  {2001})}\BibitemShut {NoStop}%
\bibitem [{\citenamefont {Lechner}\ and\ \citenamefont
  {Dellago}(2009{\natexlab{a}})}]{lechner2009defect}%
  \BibitemOpen
  \bibfield  {author} {\bibinfo {author} {\bibfnamefont {W.}~\bibnamefont
  {Lechner}}\ and\ \bibinfo {author} {\bibfnamefont {C.}~\bibnamefont
  {Dellago}},\ }\href@noop {} {\bibfield  {journal} {\bibinfo  {journal} {Soft
  Matter}\ }\textbf {\bibinfo {volume} {5}},\ \bibinfo {pages} {2752} (\bibinfo
  {year} {2009}{\natexlab{a}})}\BibitemShut {NoStop}%
\bibitem [{\citenamefont {Lechner}\ and\ \citenamefont
  {Dellago}(2009{\natexlab{b}})}]{lechner2009point}%
  \BibitemOpen
  \bibfield  {author} {\bibinfo {author} {\bibfnamefont {W.}~\bibnamefont
  {Lechner}}\ and\ \bibinfo {author} {\bibfnamefont {C.}~\bibnamefont
  {Dellago}},\ }\href@noop {} {\bibfield  {journal} {\bibinfo  {journal} {Soft
  Matter}\ }\textbf {\bibinfo {volume} {5}},\ \bibinfo {pages} {646} (\bibinfo
  {year} {2009}{\natexlab{b}})}\BibitemShut {NoStop}%
\bibitem [{\citenamefont {Dinsmore}\ \emph {et~al.}(1998)\citenamefont
  {Dinsmore}, \citenamefont {Crocker},\ and\ \citenamefont
  {Yodh}}]{DINSMORE19985}%
  \BibitemOpen
  \bibfield  {author} {\bibinfo {author} {\bibfnamefont {A.~D.}\ \bibnamefont
  {Dinsmore}}, \bibinfo {author} {\bibfnamefont {J.~C.}\ \bibnamefont
  {Crocker}},\ and\ \bibinfo {author} {\bibfnamefont {A.~G.}\ \bibnamefont
  {Yodh}},\ }\href@noop {} {\bibfield  {journal} {\bibinfo  {journal} {Current
  Opinion in Colloid \& Interface Science}\ }\textbf {\bibinfo {volume} {3}},\
  \bibinfo {pages} {5} (\bibinfo {year} {1998})}\BibitemShut {NoStop}%
\bibitem [{\citenamefont {S.~Schmitz}(1999)}]{Schmitz1999}%
  \BibitemOpen
  \bibfield  {author} {\bibinfo {author} {\bibfnamefont {K.}~\bibnamefont
  {S.~Schmitz}},\ }\href@noop {} {\bibfield  {journal} {\bibinfo  {journal}
  {Phys. Chem. Chem. Phys.}\ }\textbf {\bibinfo {volume} {1}},\ \bibinfo
  {pages} {2109} (\bibinfo {year} {1999})}\BibitemShut {NoStop}%
\bibitem [{\citenamefont {Lin}\ and\ \citenamefont {Olvera de~la
  Cruz}(2022)}]{lin2022superionic}%
  \BibitemOpen
  \bibfield  {author} {\bibinfo {author} {\bibfnamefont {Y.}~\bibnamefont
  {Lin}}\ and\ \bibinfo {author} {\bibfnamefont {M.}~\bibnamefont {Olvera de~la
  Cruz}},\ }\href@noop {} {\bibfield  {journal} {\bibinfo  {journal} {The
  Journal of Physical Chemistry B}\ }\textbf {\bibinfo {volume} {126}},\
  \bibinfo {pages} {6740} (\bibinfo {year} {2022})}\BibitemShut {NoStop}%
\bibitem [{\citenamefont {Podgornik}\ and\ \citenamefont
  {Saslow}(2005)}]{Saslow2005}%
  \BibitemOpen
  \bibfield  {author} {\bibinfo {author} {\bibfnamefont {R.}~\bibnamefont
  {Podgornik}}\ and\ \bibinfo {author} {\bibfnamefont {W.~M.}\ \bibnamefont
  {Saslow}},\ }\href {https://doi.org/10.1063/1.1908870} {\bibfield  {journal}
  {\bibinfo  {journal} {The Journal of Chemical Physics}\ }\textbf {\bibinfo
  {volume} {122}},\ \bibinfo {pages} {204902} (\bibinfo {year}
  {2005})}\BibitemShut {NoStop}%
\bibitem [{\citenamefont {Markovich}\ \emph {et~al.}(2021)\citenamefont
  {Markovich}, \citenamefont {Andelman},\ and\ \citenamefont
  {Podgornik}}]{Mar21}%
  \BibitemOpen
  \bibfield  {author} {\bibinfo {author} {\bibfnamefont {T.}~\bibnamefont
  {Markovich}}, \bibinfo {author} {\bibfnamefont {D.}~\bibnamefont
  {Andelman}},\ and\ \bibinfo {author} {\bibfnamefont {R.}~\bibnamefont
  {Podgornik}},\ }in\ \href@noop {} {\emph {\bibinfo {booktitle} {Handbook of
  Lipid Membranes}}},\ \bibinfo {editor} {edited by\ \bibinfo {editor}
  {\bibfnamefont {C.~R.}\ \bibnamefont {Safynia}}\ and\ \bibinfo {editor}
  {\bibfnamefont {J.~O.}\ \bibnamefont {Raedler}}}\ (\bibinfo  {publisher}
  {Taylor \& Francis, London},\ \bibinfo {year} {2021})\BibitemShut {NoStop}%
\bibitem [{\citenamefont {Boroudjerdi}\ \emph {et~al.}(2005)\citenamefont
  {Boroudjerdi}, \citenamefont {Kim}, \citenamefont {Naji}, \citenamefont
  {Netz}, \citenamefont {Schlagberger},\ and\ \citenamefont
  {Serr}}]{BOROUDJERDI2005129}%
  \BibitemOpen
  \bibfield  {author} {\bibinfo {author} {\bibfnamefont {H.}~\bibnamefont
  {Boroudjerdi}}, \bibinfo {author} {\bibfnamefont {Y.-W.}\ \bibnamefont
  {Kim}}, \bibinfo {author} {\bibfnamefont {A.}~\bibnamefont {Naji}}, \bibinfo
  {author} {\bibfnamefont {R.}~\bibnamefont {Netz}}, \bibinfo {author}
  {\bibfnamefont {X.}~\bibnamefont {Schlagberger}},\ and\ \bibinfo {author}
  {\bibfnamefont {A.}~\bibnamefont {Serr}},\ }\href@noop {} {\bibfield
  {journal} {\bibinfo  {journal} {Physics Reports}\ }\textbf {\bibinfo {volume}
  {416}},\ \bibinfo {pages} {129} (\bibinfo {year} {2005})}\BibitemShut
  {NoStop}%
\bibitem [{\citenamefont {Smith}\ \emph {et~al.}(2018)\citenamefont {Smith},
  \citenamefont {Maroni},\ and\ \citenamefont {Borkovec}}]{Borkovec20181}%
  \BibitemOpen
  \bibfield  {author} {\bibinfo {author} {\bibfnamefont {A.~M.}\ \bibnamefont
  {Smith}}, \bibinfo {author} {\bibfnamefont {P.}~\bibnamefont {Maroni}},\ and\
  \bibinfo {author} {\bibfnamefont {M.}~\bibnamefont {Borkovec}},\ }\href@noop
  {} {\bibfield  {journal} {\bibinfo  {journal} {Phys. Chem. Chem. Phys.}\
  }\textbf {\bibinfo {volume} {20}},\ \bibinfo {pages} {158} (\bibinfo {year}
  {2018})}\BibitemShut {NoStop}%
\bibitem [{\citenamefont {Ludwig}\ and\ \citenamefont {{von
  Klitzing}}(2020)}]{LUDWIG2020137}%
  \BibitemOpen
  \bibfield  {author} {\bibinfo {author} {\bibfnamefont {M.}~\bibnamefont
  {Ludwig}}\ and\ \bibinfo {author} {\bibfnamefont {R.}~\bibnamefont {{von
  Klitzing}}},\ }\href@noop {} {\bibfield  {journal} {\bibinfo  {journal}
  {Current Opinion in Colloid \& Interface Science}\ }\textbf {\bibinfo
  {volume} {47}},\ \bibinfo {pages} {137} (\bibinfo {year} {2020})},\ \bibinfo
  {note} {surface Forces}\BibitemShut {NoStop}%
\bibitem [{\citenamefont {Naji}\ \emph {et~al.}(2013)\citenamefont {Naji},
  \citenamefont {Kanduč}, \citenamefont {Forsman},\ and\ \citenamefont
  {Podgornik}}]{Perspective}%
  \BibitemOpen
  \bibfield  {author} {\bibinfo {author} {\bibfnamefont {A.}~\bibnamefont
  {Naji}}, \bibinfo {author} {\bibfnamefont {M.}~\bibnamefont {Kanduč}},
  \bibinfo {author} {\bibfnamefont {J.}~\bibnamefont {Forsman}},\ and\ \bibinfo
  {author} {\bibfnamefont {R.}~\bibnamefont {Podgornik}},\ }\href@noop {}
  {\bibfield  {journal} {\bibinfo  {journal} {The Journal of Chemical Physics}\
  }\textbf {\bibinfo {volume} {139}},\ \bibinfo {pages} {150901} (\bibinfo
  {year} {2013})}\BibitemShut {NoStop}%
\bibitem [{\citenamefont {Podgornik}\ and\ \citenamefont
  {Andelman}(2020)}]{podgornik2021embarras}%
  \BibitemOpen
  \bibfield  {author} {\bibinfo {author} {\bibfnamefont {R.}~\bibnamefont
  {Podgornik}}\ and\ \bibinfo {author} {\bibfnamefont {D.}~\bibnamefont
  {Andelman}},\ }\bibfield  {journal} {\bibinfo  {journal} {Journal Club for
  Condensed Matter Physics}\ }\href
  {https://doi.org/10.36471/JCCMseptember202002} {10.36471/JCCMseptember202002}
  (\bibinfo {year} {2020})\BibitemShut {NoStop}%
\bibitem [{\citenamefont {Trefalt}\ \emph {et~al.}(2017)\citenamefont
  {Trefalt}, \citenamefont {Palberg},\ and\ \citenamefont
  {Borkovec}}]{TREFALT20179}%
  \BibitemOpen
  \bibfield  {author} {\bibinfo {author} {\bibfnamefont {G.}~\bibnamefont
  {Trefalt}}, \bibinfo {author} {\bibfnamefont {T.}~\bibnamefont {Palberg}},\
  and\ \bibinfo {author} {\bibfnamefont {M.}~\bibnamefont {Borkovec}},\
  }\href@noop {} {\bibfield  {journal} {\bibinfo  {journal} {Current Opinion in
  Colloid \& Interface Science}\ }\textbf {\bibinfo {volume} {27}},\ \bibinfo
  {pages} {9} (\bibinfo {year} {2017})}\BibitemShut {NoStop}%
\bibitem [{\citenamefont {Moazzami-Gudarzi}\ \emph {et~al.}(2018)\citenamefont
  {Moazzami-Gudarzi}, \citenamefont {Adam}, \citenamefont {Smith},
  \citenamefont {Trefalt}, \citenamefont {Szilágyi}, \citenamefont {Maroni},\
  and\ \citenamefont {Borkovec}}]{Borkovec20182}%
  \BibitemOpen
  \bibfield  {author} {\bibinfo {author} {\bibfnamefont {M.}~\bibnamefont
  {Moazzami-Gudarzi}}, \bibinfo {author} {\bibfnamefont {P.}~\bibnamefont
  {Adam}}, \bibinfo {author} {\bibfnamefont {A.~M.}\ \bibnamefont {Smith}},
  \bibinfo {author} {\bibfnamefont {G.}~\bibnamefont {Trefalt}}, \bibinfo
  {author} {\bibfnamefont {I.}~\bibnamefont {Szilágyi}}, \bibinfo {author}
  {\bibfnamefont {P.}~\bibnamefont {Maroni}},\ and\ \bibinfo {author}
  {\bibfnamefont {M.}~\bibnamefont {Borkovec}},\ }\href@noop {} {\bibfield
  {journal} {\bibinfo  {journal} {Phys. Chem. Chem. Phys.}\ }\textbf {\bibinfo
  {volume} {20}},\ \bibinfo {pages} {9436} (\bibinfo {year}
  {2018})}\BibitemShut {NoStop}%
\bibitem [{\citenamefont {Sutton}(2020)}]{Sutton}%
  \BibitemOpen
  \bibfield  {author} {\bibinfo {author} {\bibfnamefont {A.~P.}\ \bibnamefont
  {Sutton}},\ }\href@noop {} {\emph {\bibinfo {title} {{Physics of Elasticity
  and Crystal Defects}}}}\ (\bibinfo  {publisher} {Oxford University Press,
  Oxford},\ \bibinfo {year} {2020})\BibitemShut {NoStop}%
\bibitem [{\citenamefont {Lechner}\ \emph {et~al.}(2013)\citenamefont
  {Lechner}, \citenamefont {Polster}, \citenamefont {Maret}, \citenamefont
  {Keim},\ and\ \citenamefont {Dellago}}]{lechner2013self}%
  \BibitemOpen
  \bibfield  {author} {\bibinfo {author} {\bibfnamefont {W.}~\bibnamefont
  {Lechner}}, \bibinfo {author} {\bibfnamefont {D.}~\bibnamefont {Polster}},
  \bibinfo {author} {\bibfnamefont {G.}~\bibnamefont {Maret}}, \bibinfo
  {author} {\bibfnamefont {P.}~\bibnamefont {Keim}},\ and\ \bibinfo {author}
  {\bibfnamefont {C.}~\bibnamefont {Dellago}},\ }\href@noop {} {\bibfield
  {journal} {\bibinfo  {journal} {Physical Review E}\ }\textbf {\bibinfo
  {volume} {88}},\ \bibinfo {pages} {060402} (\bibinfo {year}
  {2013})}\BibitemShut {NoStop}%
\bibitem [{\citenamefont {Eisenmann}\ \emph {et~al.}(2005)\citenamefont
  {Eisenmann}, \citenamefont {Gasser}, \citenamefont {Keim}, \citenamefont
  {Maret},\ and\ \citenamefont {von Gr{\"u}nberg}}]{eisenmann2005pair}%
  \BibitemOpen
  \bibfield  {author} {\bibinfo {author} {\bibfnamefont {C.}~\bibnamefont
  {Eisenmann}}, \bibinfo {author} {\bibfnamefont {U.}~\bibnamefont {Gasser}},
  \bibinfo {author} {\bibfnamefont {P.}~\bibnamefont {Keim}}, \bibinfo {author}
  {\bibfnamefont {G.}~\bibnamefont {Maret}},\ and\ \bibinfo {author}
  {\bibfnamefont {H.-H.}\ \bibnamefont {von Gr{\"u}nberg}},\ }\href@noop {}
  {\bibfield  {journal} {\bibinfo  {journal} {Physical review letters}\
  }\textbf {\bibinfo {volume} {95}},\ \bibinfo {pages} {185502} (\bibinfo
  {year} {2005})}\BibitemShut {NoStop}%
\bibitem [{\citenamefont {He}\ and\ \citenamefont
  {Chen}(2013)}]{he2013interaction}%
  \BibitemOpen
  \bibfield  {author} {\bibinfo {author} {\bibfnamefont {B.}~\bibnamefont
  {He}}\ and\ \bibinfo {author} {\bibfnamefont {Y.}~\bibnamefont {Chen}},\
  }\href@noop {} {\bibfield  {journal} {\bibinfo  {journal} {Solid State
  Communications}\ }\textbf {\bibinfo {volume} {159}},\ \bibinfo {pages} {60}
  (\bibinfo {year} {2013})}\BibitemShut {NoStop}%
\bibitem [{\citenamefont {Pertsinidis}\ and\ \citenamefont
  {Ling}(2001)}]{pertsinidis2001equilibrium}%
  \BibitemOpen
  \bibfield  {author} {\bibinfo {author} {\bibfnamefont {A.}~\bibnamefont
  {Pertsinidis}}\ and\ \bibinfo {author} {\bibfnamefont {X.}~\bibnamefont
  {Ling}},\ }\href@noop {} {\bibfield  {journal} {\bibinfo  {journal} {Physical
  Review Letters}\ }\textbf {\bibinfo {volume} {87}},\ \bibinfo {pages}
  {098303} (\bibinfo {year} {2001})}\BibitemShut {NoStop}%
\bibitem [{\citenamefont {Lechner}\ \emph {et~al.}(2014)\citenamefont
  {Lechner}, \citenamefont {B{\"u}chler},\ and\ \citenamefont
  {Zoller}}]{lechner2014role}%
  \BibitemOpen
  \bibfield  {author} {\bibinfo {author} {\bibfnamefont {W.}~\bibnamefont
  {Lechner}}, \bibinfo {author} {\bibfnamefont {H.-P.}\ \bibnamefont
  {B{\"u}chler}},\ and\ \bibinfo {author} {\bibfnamefont {P.}~\bibnamefont
  {Zoller}},\ }\href@noop {} {\bibfield  {journal} {\bibinfo  {journal}
  {Physical Review Letters}\ }\textbf {\bibinfo {volume} {112}},\ \bibinfo
  {pages} {255301} (\bibinfo {year} {2014})}\BibitemShut {NoStop}%
\bibitem [{\citenamefont {Bardeen}\ \emph {et~al.}(1957)\citenamefont
  {Bardeen}, \citenamefont {Cooper},\ and\ \citenamefont
  {Schrieffer}}]{bardeen1957theory}%
  \BibitemOpen
  \bibfield  {author} {\bibinfo {author} {\bibfnamefont {J.}~\bibnamefont
  {Bardeen}}, \bibinfo {author} {\bibfnamefont {L.~N.}\ \bibnamefont
  {Cooper}},\ and\ \bibinfo {author} {\bibfnamefont {J.~R.}\ \bibnamefont
  {Schrieffer}},\ }\href@noop {} {\bibfield  {journal} {\bibinfo  {journal}
  {Physical review}\ }\textbf {\bibinfo {volume} {108}},\ \bibinfo {pages}
  {1175} (\bibinfo {year} {1957})}\BibitemShut {NoStop}%
\bibitem [{\citenamefont {Wu}\ \emph {et~al.}(2024)\citenamefont {Wu},
  \citenamefont {Ou-Yang},\ and\ \citenamefont
  {Podgornik}}]{wu2024electrostatic}%
  \BibitemOpen
  \bibfield  {author} {\bibinfo {author} {\bibfnamefont {H.}~\bibnamefont
  {Wu}}, \bibinfo {author} {\bibfnamefont {Z.-C.}\ \bibnamefont {Ou-Yang}},\
  and\ \bibinfo {author} {\bibfnamefont {R.}~\bibnamefont {Podgornik}},\
  }\href@noop {} {\bibfield  {journal} {\bibinfo  {journal} {Europhysics
  Letters}\ }\textbf {\bibinfo {volume} {148}},\ \bibinfo {pages} {47001}
  (\bibinfo {year} {2024})}\BibitemShut {NoStop}%
\bibitem [{\citenamefont {Xie}\ and\ \citenamefont
  {Huang}(1994)}]{QIanXie1994}%
  \BibitemOpen
  \bibfield  {author} {\bibinfo {author} {\bibfnamefont {Q.}~\bibnamefont
  {Xie}}\ and\ \bibinfo {author} {\bibfnamefont {M.}~\bibnamefont {Huang}},\
  }\href@noop {} {\bibfield  {journal} {\bibinfo  {journal} {physica status
  solidi (b)}\ }\textbf {\bibinfo {volume} {186}},\ \bibinfo {pages} {393}
  (\bibinfo {year} {1994})}\BibitemShut {NoStop}%
\bibitem [{\citenamefont {L{\"o}wen}\ \emph {et~al.}(1993)\citenamefont
  {L{\"o}wen}, \citenamefont {Hansen},\ and\ \citenamefont
  {Madden}}]{Lowen1993}%
  \BibitemOpen
  \bibfield  {author} {\bibinfo {author} {\bibfnamefont {H.}~\bibnamefont
  {L{\"o}wen}}, \bibinfo {author} {\bibfnamefont {J.-P.}\ \bibnamefont
  {Hansen}},\ and\ \bibinfo {author} {\bibfnamefont {P.~A.}\ \bibnamefont
  {Madden}},\ }\href@noop {} {\bibfield  {journal} {\bibinfo  {journal} {The
  Journal of chemical physics}\ }\textbf {\bibinfo {volume} {98}},\ \bibinfo
  {pages} {3275} (\bibinfo {year} {1993})}\BibitemShut {NoStop}%
\bibitem [{\citenamefont {Maggs}\ and\ \citenamefont
  {Podgornik}(2016)}]{maggs2016general}%
  \BibitemOpen
  \bibfield  {author} {\bibinfo {author} {\bibfnamefont {A.}~\bibnamefont
  {Maggs}}\ and\ \bibinfo {author} {\bibfnamefont {R.}~\bibnamefont
  {Podgornik}},\ }\href@noop {} {\bibfield  {journal} {\bibinfo  {journal}
  {Soft matter}\ }\textbf {\bibinfo {volume} {12}},\ \bibinfo {pages} {1219}
  (\bibinfo {year} {2016})}\BibitemShut {NoStop}%
\bibitem [{\citenamefont {Schwerdtfeger}\ \emph {et~al.}(2021)\citenamefont
  {Schwerdtfeger}, \citenamefont {Burrows},\ and\ \citenamefont
  {Smits}}]{Schwerdtfeger2021}%
  \BibitemOpen
  \bibfield  {author} {\bibinfo {author} {\bibfnamefont {P.}~\bibnamefont
  {Schwerdtfeger}}, \bibinfo {author} {\bibfnamefont {A.}~\bibnamefont
  {Burrows}},\ and\ \bibinfo {author} {\bibfnamefont {O.~R.}\ \bibnamefont
  {Smits}},\ }\href@noop {} {\bibfield  {journal} {\bibinfo  {journal} {The
  Journal of Physical Chemistry A}\ }\textbf {\bibinfo {volume} {125}},\
  \bibinfo {pages} {3037} (\bibinfo {year} {2021})}\BibitemShut {NoStop}%
\bibitem [{\citenamefont {Ashcroft}\ and\ \citenamefont
  {Mermin}(1976)}]{ashcroft1976}%
  \BibitemOpen
  \bibfield  {author} {\bibinfo {author} {\bibfnamefont {N.~W.}\ \bibnamefont
  {Ashcroft}}\ and\ \bibinfo {author} {\bibfnamefont {N.~D.}\ \bibnamefont
  {Mermin}},\ }\href@noop {} {\emph {\bibinfo {title} {Solid State Physics}}}\
  (\bibinfo  {publisher} {Saunders College Publishing, Philadelphia, PA},\
  \bibinfo {year} {1976})\BibitemShut {NoStop}%
\bibitem [{\citenamefont {Kung}(2009)}]{Kung}%
  \BibitemOpen
  \bibfield  {author} {\bibinfo {author} {\bibfnamefont {W.}~\bibnamefont
  {Kung}},\ }\href {https://doi.org/10.1142/7026} {\emph {\bibinfo {title}
  {Geometry and Phase Transitions in Colloids and Polymers}}}\ (\bibinfo
  {publisher} {World Scientific, Singapore},\ \bibinfo {year}
  {2009})\BibitemShut {NoStop}%
\bibitem [{\citenamefont {Dobnikar}\ \emph {et~al.}(2002)\citenamefont
  {Dobnikar}, \citenamefont {Chen}, \citenamefont {Rzehak},\ and\ \citenamefont
  {von Grünberg}}]{Dobnikar_2003}%
  \BibitemOpen
  \bibfield  {author} {\bibinfo {author} {\bibfnamefont {J.}~\bibnamefont
  {Dobnikar}}, \bibinfo {author} {\bibfnamefont {Y.}~\bibnamefont {Chen}},
  \bibinfo {author} {\bibfnamefont {R.}~\bibnamefont {Rzehak}},\ and\ \bibinfo
  {author} {\bibfnamefont {H.~H.}\ \bibnamefont {von Grünberg}},\ }\href
  {https://doi.org/10.1088/0953-8984/15/1/335} {\bibfield  {journal} {\bibinfo
  {journal} {Journal of Physics: Condensed Matter}\ }\textbf {\bibinfo {volume}
  {15}},\ \bibinfo {pages} {S263} (\bibinfo {year} {2002})}\BibitemShut
  {NoStop}%
\bibitem [{Note1()}]{Note1}%
  \BibitemOpen
  \bibinfo {note} {$ \mu \DOTSI \ointop \ilimits@ _{S} {\protect \bm {n}} \cdot
  \left [ {\textstyle \protect \frac 12} \protect \mbox {\protect \boldmath
  $\nabla $}{\protect \bm {u}}^2- {\protect \bm {u}} \times (\protect \mbox
  {\protect \boldmath $\nabla $}\times {\protect \bm {u}}) - {\protect \bm {u}}
  (\protect \mbox {\protect \boldmath $\nabla $}\cdot {\protect \bm {u}})\right
  ] {\protect \rm d}S $}\BibitemShut {NoStop}%
\bibitem [{\citenamefont {Teodosiu}(1982)}]{Teodosiu1982}%
  \BibitemOpen
  \bibfield  {author} {\bibinfo {author} {\bibfnamefont {C.}~\bibnamefont
  {Teodosiu}},\ }\bibinfo {title} {The elastic field of point defects},\ in\
  \href {https://doi.org/10.1007/978-3-662-11634-0_5} {\emph {\bibinfo
  {booktitle} {Elastic Models of Crystal Defects}}}\ (\bibinfo  {publisher}
  {Springer Berlin Heidelberg},\ \bibinfo {address} {Berlin, Heidelberg},\
  \bibinfo {year} {1982})\ pp.\ \bibinfo {pages} {287--316}\BibitemShut
  {NoStop}%
\bibitem [{\citenamefont {Kardar}\ and\ \citenamefont
  {Golestanian}(1999)}]{RevModPhys.71.1233}%
  \BibitemOpen
  \bibfield  {author} {\bibinfo {author} {\bibfnamefont {M.}~\bibnamefont
  {Kardar}}\ and\ \bibinfo {author} {\bibfnamefont {R.}~\bibnamefont
  {Golestanian}},\ }\href@noop {} {\bibfield  {journal} {\bibinfo  {journal}
  {Rev. Mod. Phys.}\ }\textbf {\bibinfo {volume} {71}},\ \bibinfo {pages}
  {1233} (\bibinfo {year} {1999})}\BibitemShut {NoStop}%
\bibitem [{\citenamefont {Karimi Pour~Haddadan}\ \emph
  {et~al.}(2019)\citenamefont {Karimi Pour~Haddadan}, \citenamefont {Naji},\
  and\ \citenamefont {Podgornik}}]{Karimi2019}%
  \BibitemOpen
  \bibfield  {author} {\bibinfo {author} {\bibfnamefont {F.}~\bibnamefont
  {Karimi Pour~Haddadan}}, \bibinfo {author} {\bibfnamefont {A.}~\bibnamefont
  {Naji}},\ and\ \bibinfo {author} {\bibfnamefont {R.}~\bibnamefont
  {Podgornik}},\ }\href {https://doi.org/10.1039/C8SM02328J} {\bibfield
  {journal} {\bibinfo  {journal} {Soft Matter}\ }\textbf {\bibinfo {volume}
  {15}},\ \bibinfo {pages} {2216} (\bibinfo {year} {2019})}\BibitemShut
  {NoStop}%
\bibitem [{\citenamefont {Walz}\ and\ \citenamefont
  {Fuchs}(2010)}]{PhysRevB.81.134110}%
  \BibitemOpen
  \bibfield  {author} {\bibinfo {author} {\bibfnamefont {C.}~\bibnamefont
  {Walz}}\ and\ \bibinfo {author} {\bibfnamefont {M.}~\bibnamefont {Fuchs}},\
  }\href@noop {} {\bibfield  {journal} {\bibinfo  {journal} {Phys. Rev. B}\
  }\textbf {\bibinfo {volume} {81}},\ \bibinfo {pages} {134110} (\bibinfo
  {year} {2010})}\BibitemShut {NoStop}%
\bibitem [{\citenamefont {Landau}\ and\ \citenamefont
  {Lifshitz}(1970)}]{Landau}%
  \BibitemOpen
  \bibfield  {author} {\bibinfo {author} {\bibfnamefont {L.}~\bibnamefont
  {Landau}}\ and\ \bibinfo {author} {\bibfnamefont {E.}~\bibnamefont
  {Lifshitz}},\ }\href@noop {} {\emph {\bibinfo {title} {Theory of Elasticity,
  2nd Edition, Course of Theoretical Physics, Vol. 7}}}\ (\bibinfo  {publisher}
  {Pergamon Press, Oxford},\ \bibinfo {year} {1970})\BibitemShut {NoStop}%
\bibitem [{\citenamefont {Verwey}\ and\ \citenamefont
  {Overbeek}(1948)}]{Verwey}%
  \BibitemOpen
  \bibfield  {author} {\bibinfo {author} {\bibfnamefont {E.~J.~W.}\
  \bibnamefont {Verwey}}\ and\ \bibinfo {author} {\bibfnamefont {J.~T.~G.}\
  \bibnamefont {Overbeek}},\ }\href@noop {} {\emph {\bibinfo {title} {Theory of
  the Stability of Lyophobic Colloids}}}\ (\bibinfo  {publisher} {Elsevier, New
  York},\ \bibinfo {year} {1948})\BibitemShut {NoStop}%
\bibitem [{\citenamefont {Wiegel}(1975)}]{WIEGEL197557}%
  \BibitemOpen
  \bibfield  {author} {\bibinfo {author} {\bibfnamefont {F.}~\bibnamefont
  {Wiegel}},\ }\href@noop {} {\bibfield  {journal} {\bibinfo  {journal}
  {Physics Reports}\ }\textbf {\bibinfo {volume} {16}},\ \bibinfo {pages} {57}
  (\bibinfo {year} {1975})}\BibitemShut {NoStop}%
\bibitem [{\citenamefont {Blossey}\ and\ \citenamefont
  {Maggs}(2018)}]{Blossey_2018}%
  \BibitemOpen
  \bibfield  {author} {\bibinfo {author} {\bibfnamefont {R.}~\bibnamefont
  {Blossey}}\ and\ \bibinfo {author} {\bibfnamefont {A.~C.}\ \bibnamefont
  {Maggs}},\ }\href@noop {} {\bibfield  {journal} {\bibinfo  {journal} {Journal
  of Physics A: Mathematical and Theoretical}\ }\textbf {\bibinfo {volume}
  {51}},\ \bibinfo {pages} {385001} (\bibinfo {year} {2018})}\BibitemShut
  {NoStop}%
\bibitem [{\citenamefont {Avni}\ \emph {et~al.}(2019)\citenamefont {Avni},
  \citenamefont {Andelman},\ and\ \citenamefont {Podgornik}}]{AVNI201970}%
  \BibitemOpen
  \bibfield  {author} {\bibinfo {author} {\bibfnamefont {Y.}~\bibnamefont
  {Avni}}, \bibinfo {author} {\bibfnamefont {D.}~\bibnamefont {Andelman}},\
  and\ \bibinfo {author} {\bibfnamefont {R.}~\bibnamefont {Podgornik}},\
  }\href@noop {} {\bibfield  {journal} {\bibinfo  {journal} {Current Opinion in
  Electrochemistry}\ }\textbf {\bibinfo {volume} {13}},\ \bibinfo {pages} {70}
  (\bibinfo {year} {2019})},\ \bibinfo {note} {fundamental and Theoretical
  Electrochemistry -- Physical and Nanoelectrochemistry}\BibitemShut {NoStop}%
\bibitem [{\citenamefont {Naji}\ \emph {et~al.}(2011)\citenamefont {Naji},
  \citenamefont {Kandu{\v{c}}}, \citenamefont {Netz},\ and\ \citenamefont
  {Podgornik}}]{naji2011exotic}%
  \BibitemOpen
  \bibfield  {author} {\bibinfo {author} {\bibfnamefont {A.}~\bibnamefont
  {Naji}}, \bibinfo {author} {\bibfnamefont {M.}~\bibnamefont {Kandu{\v{c}}}},
  \bibinfo {author} {\bibfnamefont {R.~R.}\ \bibnamefont {Netz}},\ and\
  \bibinfo {author} {\bibfnamefont {R.}~\bibnamefont {Podgornik}},\ }in\
  \href@noop {} {\emph {\bibinfo {booktitle} {Understanding Soft Condensed
  Matter via Modeling and Computation}}}\ (\bibinfo  {publisher} {World
  Scientific, Singapore},\ \bibinfo {year} {2011})\ pp.\ \bibinfo {pages}
  {265--295}\BibitemShut {NoStop}%
\bibitem [{\citenamefont {Liu}\ \emph {et~al.}(2022)\citenamefont {Liu},
  \citenamefont {Stenhammar}, \citenamefont {Wennerstr{\" o}m},\ and\
  \citenamefont {Sparr}}]{Liu2022}%
  \BibitemOpen
  \bibfield  {author} {\bibinfo {author} {\bibfnamefont {X.}~\bibnamefont
  {Liu}}, \bibinfo {author} {\bibfnamefont {J.}~\bibnamefont {Stenhammar}},
  \bibinfo {author} {\bibfnamefont {H.}~\bibnamefont {Wennerstr{\" o}m}},\ and\
  \bibinfo {author} {\bibfnamefont {E.}~\bibnamefont {Sparr}},\ }\href@noop {}
  {\bibfield  {journal} {\bibinfo  {journal} {The Journal of Physical Chemistry
  Letters}\ }\textbf {\bibinfo {volume} {13}},\ \bibinfo {pages} {498}
  (\bibinfo {year} {2022})}\BibitemShut {NoStop}%
\bibitem [{\citenamefont {Kornyshev}(2007)}]{Kornyshev2007}%
  \BibitemOpen
  \bibfield  {author} {\bibinfo {author} {\bibfnamefont {A.~A.}\ \bibnamefont
  {Kornyshev}},\ }\href@noop {} {\bibfield  {journal} {\bibinfo  {journal} {The
  Journal of Physical Chemistry B}\ }\textbf {\bibinfo {volume} {111}},\
  \bibinfo {pages} {5545} (\bibinfo {year} {2007})}\BibitemShut {NoStop}%
\end{thebibliography}%

\end{document}